# Brewster-anomaly delocalization for free-electron radiation

Zheng Gong,[1,2] Xiangfeng Xi,[1] Hongsheng Chen,[1] Owen D. Miller,[3,*]
Stefan Rotter,[2,*] and Xiao Lin[1,*]

**Corresponding authors: xiaolinzju@zju.edu.cn (X. Lin); stefan.rotter@tuwien.ac.at (S. Rotter); owen.miller@yale.edu (O. D. Miller).*

[1]*State Key Laboratory of Extreme Photonics and Instrumentation, Zhejiang Key Laboratory of Intelligent Electromagnetic Control and Advanced Electronic Integration, College of Information Science & Electronic Engineering, Zhejiang University, Hangzhou 310027, China.*

[2]*Institute for Theoretical Physics, Vienna University of Technology (TU Wien), Vienna A-1040, Austria.*

[3]*Department of Applied Physics, Yale University, New Haven, Connecticut 06520, USA.*

Localization effects are central to disordered electronics and photonics. In electronics, Anderson localization governs electron confinement in randomly perturbed lattices. Similarly, its photonic counterpart inhibits light transport via disorder – but with a unique exception: Brewster-anomaly delocalization, where the Brewster effect prevents multiple-scattering interference and counteracts the localization. Despite extensive research in electronics and photonics separately, the intricate role of localization effects in free-electron–light interactions – vital for lasers, accelerators, microscopy and spectroscopy, and quantum information – remains largely unexplored. At the same time, localization effects are widely regarded as a key factor limiting the efficient coupling between free electrons and light in random media. Here we overcome this key limitation via the unconventional interplay between Brewster-anomaly delocalization and free-electron radiation. In this way, free-electron radiation can be localization-free, intense and directional even in strongly disordered, unengineered multilayers. Essentially, this delocalization-mediated free-electron radiation is remarkably invariant not only to the random-medium configuration, but also to the light frequency and the electron velocity. Our findings unlock new opportunities for particle detectors and achromatic light sources operating in easy-to-fabricate complex media at previously inaccessible frequencies.

The interaction between free charged particles and matter constitutes a basic pathway to light emission known as charged-particle radiation, for which free-electron radiation [1–8] is a prominent example. The intensity or directionality of this radiation is precisely correlated with the particle kinetic energy or velocity, forming a key principle underpinning particle detectors [9–12]. These detectors not only have facilitated a golden age of discovering new particles [13–15], but also continue the "party of high energy physics" [16] by searching for deviations from the Standard Model of elementary particles [17–21]. Beyond its established role in particle physics, charged-particle radiation has emerged as a versatile resource in nanophotonics. By granting access to light emission at virtually any frequency, charged-particle radiation empowers a broad spectrum of applications, ranging from tailored light sources [22–25], free-electron lasers [26–28], accelerators [29,30] and magnonic devices [31], to advanced microscopy and spectroscopy [32–35], as well as quantum information technologies [8,36,37].

Central to these advances in particle physics and nanophotonics driven by charged-particle radiation is a paradigm which heavily hinges on modelling free charged-particle interactions with artificial nanostructures. Within this paradigm, structural order typically constitutes a key ingredient. A prime example is free-electron radiation in systems with intrinsic periodicity [38–45] or hidden long-range order [46–48], such as metamaterials, photonic crystals, and photonic quasicrystals [49–52]. Recently, Ref. [53] suggested an alternative route to manipulate free-electron radiation via *judiciously engineered* resonances in pseudo-random media. However, the prohibitive fabrication costs, limited scalability, and stringent tolerance requirements are inherent to structurally ordered designs and engineered resonances. These drawbacks strongly motivate the exploration of free-electron radiation in *unengineered* random media – systems that enable low-cost, large-scale, and imperfection-robust fabrication.

Nevertheless, operating in the regime of *strong*, *unengineered* randomness poses a central challenge. The resulting properties (e.g., intensity and directionality) of free-electron radiation are typically regarded as intrinsically random and therefore difficult to predict or control. This apparent link between structural disorder and uncontrolled emission raises a core question: how can free-electron radiation be harnessed in strongly disordered media without relying on judicious structural design? From the viewpoint of Anderson localization [54–59], this challenge appears particularly severe, since strong disorder suppresses light transport entirely. As a result, the emitted radiation tends to become localized inside structures with strong disorder, leading to intrinsically low light extraction efficiency. In essence, localization effects impose a key limitation on the effective coupling between free electrons and radiated light in random media.

Here, we overcome this key limitation by exploiting Brewster-anomaly delocalization, which is a unique exception to Anderson localization. Regardless of the structural randomness, the effect of Brewster-anomaly delocalization dictates that the localization length (i.e., the critical length of layered random media required for localization) of $p$-polarized light diverges at the Brewster angle [60,61]. Ever since its theoretical and experimental identification [60–64], Brewster-anomaly delocalization has been widely studied in random systems incorporating metamaterial composites [65,66], correlated disorder hosting new localization mechanisms beyond the Anderson model [67–69], and emerging time-varying media [70]. These studies have far-reaching implications across diverse areas extending even to biological optics [71,72]. Despite these advances and a long research history in wave control in random media [73–77], Brewster-anomaly delocalization has never been connected to the framework of free-electron radiation. In particular, whether Brewster-anomaly delocalization could impart robust features to free-electron radiation in the presence of strong, unengineered disorder and/or enable its efficient harnessing in random media, remains unknown.

Here we show that free-electron radiation in strongly disordered, unengineered random media can be rendered both intense and directional. On the one hand, Brewster-anomaly delocalization creates intense free-electron radiation by freeing it from otherwise unavoidable localization. On the other hand, the localization-to-delocalization transition from the non-Brewster to the Brewster angle yields directional emission precisely at the Brewster angle. Moreover, rooted in the statistically *invariant* nature of Brewster-anomaly delocalization, the delocalized free-electron radiation we report is robust not only against unengineered variations in the random medium, but also against variations in light frequency and particle velocity. These robust features set our findings apart from *all* known forms of free-electron radiation from random media [46–48,53], such as those based on long-range order or engineered resonances. In short, our findings capture the best of both worlds of free-electron radiation and localization/delocalization effects, yielding intense and directional free-electron emission while maintaining the delocalization-enabled robustness.

Figure 1 provides a conceptual illustration of Brewster-anomaly delocalization for free-electron radiation in strongly disordered, unengineered random media. An electron moving at constant speed penetrates a layered random nanostructure composed of alternating dielectrics I and II [Fig. 1(a)]. Without loss of generality, both dielectrics are assumed to be non-dispersive. The unengineered structural randomness is encoded in the thickness of each layer, which is a continuous random variable governed by a probability density function $\wp\left(d_0, \sigma^2\right)$ (e.g., an exponential distribution), with $d_0$ and $\sigma$ denoting the mean value and standard deviation of this distribution, respectively. Physically, $\sigma/\lambda_0$ quantifies the disorder strength, where $\lambda_0$ is the working wavelength of light in vacuum. In the weak-disorder regime (i.e., $\sigma/\lambda_0 \ll 1$), $d_0/\lambda_0$ further characterizes the random medium on average as metamaterial-like (i.e., $d_0/\lambda_0 \ll 1$) or as photonic-crystal-like (i.e., $d_0/\lambda_0 \cong 1$) [78]. In this work, both $d_0/\lambda_0$ and $\sigma/\lambda_0$ are allowed to take arbitrary positive values, placing free-electron radiation in the most general and least predictable regime.

In principle, the resulting free-electron radiation (i.e., transition radiation) is fully coherent, since the quantum-mechanical state of the sample remains unchanged by the electron's penetration [2,5,37,79]. To intuitively appreciate the stochastic nature of free-electron radiation from the random medium, its origin can be modeled as a random array of complex dipoles, all of which are mutually coherent. Specifically, across the broad free-electron radiation spectrum (e.g., from microwave to X-ray), each interface penetrated by the impinging electron behaves as a *p*-polarized vertical electric dipole at the frequency of interest. The angular pattern of each dipole is determined by the particle velocity and the surrounding media [79–82]. The strong, unengineered disorder then results in a random distribution of the initial dipole phases, in contrast to cases with hidden order or pseudo-randomness [46–48,53].

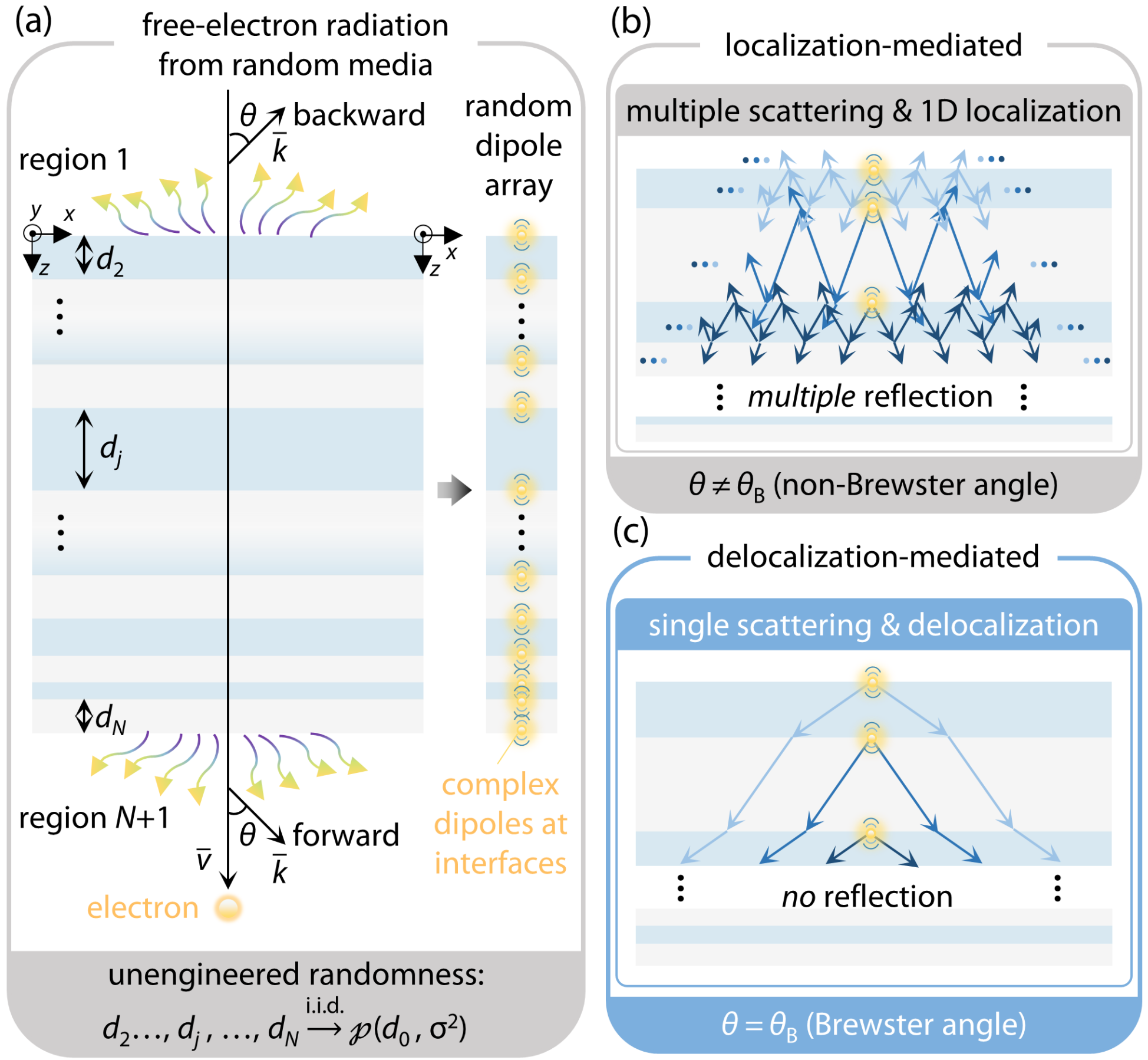

FIG. 1. Conceptual illustration of free-electron radiation from random media mediated by Brewster-anomaly delocalization. (a) Structural schematic. A uniformly moving electron with velocity $\overline{v} = \hat{z}v$ travels perpendicularly through a random multilayer consisting of alternating dielectrics I and II. The layer thicknesses $d_j$ $(j = 2 \ldots N)$ are independent and identically distributed (i.i.d.) *continuous* random variables with the probability density function $\wp(d_0, \sigma^2)$, where $d_0$ and $\sigma$ denote the mean value and standard deviation, respectively. As the electron penetrates the random medium, $p$-polarized light is emitted at each interface. For conceptual simplicity, the total free-electron radiation at the frequency of interest can be modeled as a random array of vertical electric dipoles positioned at each interface, with angular radiation patterns determined by the dielectric relative permittivities ($\varepsilon_{\mathrm{r,I}}$ and $\varepsilon_{\mathrm{r,II}}$) and the electron velocity $v$. The radiation angle $\theta$ is defined as the angle between the wavevector of light and the electron velocity (or the reversed electron velocity) in the substrate region $N+1$ (or the superstrate region 1). (b),(c) Free-electron radiation in random media mediated by conventional localization effects and Brewster-anomaly delocalization. Unless otherwise specified, throughout the main text we use $\varepsilon_{\mathrm{I}} = 1$, $\varepsilon_{\mathrm{II}} = 2.1$, $v = v_0 = 0.5c < \min\left(c/\sqrt{\varepsilon_{\mathrm{I}}}, c/\sqrt{\varepsilon_{\mathrm{II}}}\right)$, a total number of interfaces $N = 5 \times 10^4$,

working wavelength of light in vacuum $\lambda_0 = 500$ nm, and the thickness of each layer $d_j$ follows an exponential probability density function $\mathcal{p}(d_0, \sigma^2) = e^{-d_j/d_0}/d_0$, where $\sigma = d_0 = \lambda_0$.

Generally, free-electron radiation is mediated and limited by localization effects [Fig. 1(b)]. As background, localization of light in layered random media is a well-established phenomenon, as demonstrated by numerous experimental studies [57,62–64,83–85]. Following the spirit of light localization and extending it to free-electron–light interactions, we decompose the free-electron radiation into plane-wave components characterized by a fixed transverse wavevector (i.e., a fixed emission angle). Each such component propagates through the layered structure as an effectively one-dimensional system, undergoing repeated specular reflections at the interfaces. In the presence of strong disorder, these reflections give rise to multiple scattering whose cumulative effect suppresses transmission and *localizes* free-electron radiation within the structure, limiting the far-field light extraction efficiency.

In stark contrast, free-electron radiation mediated by Brewster-anomaly delocalization can overcome this limitation [Fig. 1(c)]. At the Brewster angle, radiation emitted at $\theta = \theta_\mathrm{B}$ undergoes vanishing reflection at each interface (whether passing from dielectric I to II or vice versa). As a result, the repeated backscattering processes required for localization are suppressed. Consequently, free-electron radiation at the Brewster angle can be effectively modeled as *delocalized* even in the presence of sufficiently strong, unengineered disorder. At this point, a pathway toward realizing Brewster-anomaly delocalization in practice and harnessing it for free-electron radiation is established. First, there is a natural compatibility between Brewster-anomaly delocalization and free-electron radiation, as both can be specific to *p*-polarized light. Second, owing to the unengineered nature of the system, we simply require sufficient disorder strength, a large number of interfaces, and a proper index contrast for the binary multilayer stack to induce the localization-to-delocalization transition.

We now proceed to show that intense and directional free-electron radiation indeed emerges from random media via Brewster-anomaly delocalization in Fig. 2. The radiation phenomena are classified into three representative regimes [Fig. 2(a)]. The first regime corresponds to conventional localization-mediated free-electron radiation away from the Brewster angle (e.g., $\theta \approx 1.2\theta_{\mathrm{B}}$). When the number of interfaces $N$ is relatively small, the configurationally averaged angular spectral energy density $\langle U \rangle_{\mathrm{g}}$ grows linearly with $N$, where $\langle \cdot \rangle_{\mathrm{g}}$ represents the geometric average over multiple random configurations. Free-electron radiation emitted in both forward and backward directions (i.e., relative to the direction of electron motion) is summed in $\langle U \rangle_{\mathrm{g}}$. The angular spectral energy density $U$ is numerically obtained by extending Ginzburg and Frank's theory [79,80] [see also supplementary section S1]. The revealed linear growth of $\langle U \rangle_{\mathrm{g}} \sim N$ is consistent with the canonical picture of X-ray transition-radiation particle detectors operated in the weak-absorption regime [10,12]. Yet, when the number of interfaces becomes sufficiently large, localization effects begin to dominate, leading to a pronounced slowdown of the intensity growth and ultimately driving it to saturation. The saturation indicates the balance between the localization-induced suppression and the enhancement originating from an increased number of radiators (i.e., interfaces). In the second regime where the radiation angle is near, but not exactly at the Brewster angle (e.g., $\theta \approx 0.9\theta_{\mathrm{B}}$), the radiation phenomenon shows a similar trend, except that a larger number of interfaces is required to reach the saturation. The larger critical number of interfaces owes its origin to the increased localization length [see also Fig. S1]. For the third regime exactly at the Brewster angle (i.e., $\theta = \theta_{\mathrm{B}}$), the localization length is further increased and divergent, meaning that the localization effect is eliminated entirely. In this scenario, regardless of the number of interfaces, the radiation intensity grows linearly at the Brewster angle, confirming the emergence of intense free-electron radiation from strongly disordered, unengineered random systems.

Putting all three regimes together, we further show in Figs. 2(b)–2(d) that free-electron radiation from strongly disordered, unengineered random media can be highly directional owing to the contrasting mechanisms governing free-electron radiation at the non-Brewster and Brewster angles. In fact, the system size required for delocalized, directional free-electron radiation can be quantitatively estimated from the localization length near the Brewster angle [see also Fig. S2]. As the number of interfaces increases from Fig. 2(b) to Fig. 2(d), the radiation pattern evolves from a relatively broad and random profile into directional emission with a sharp radiation peak at the Brewster angle. This intense and directional free-electron radiation persists for various disorder distributions [see also Fig. S3], indicating that our finding is robust against the specific choice of disorder distribution.

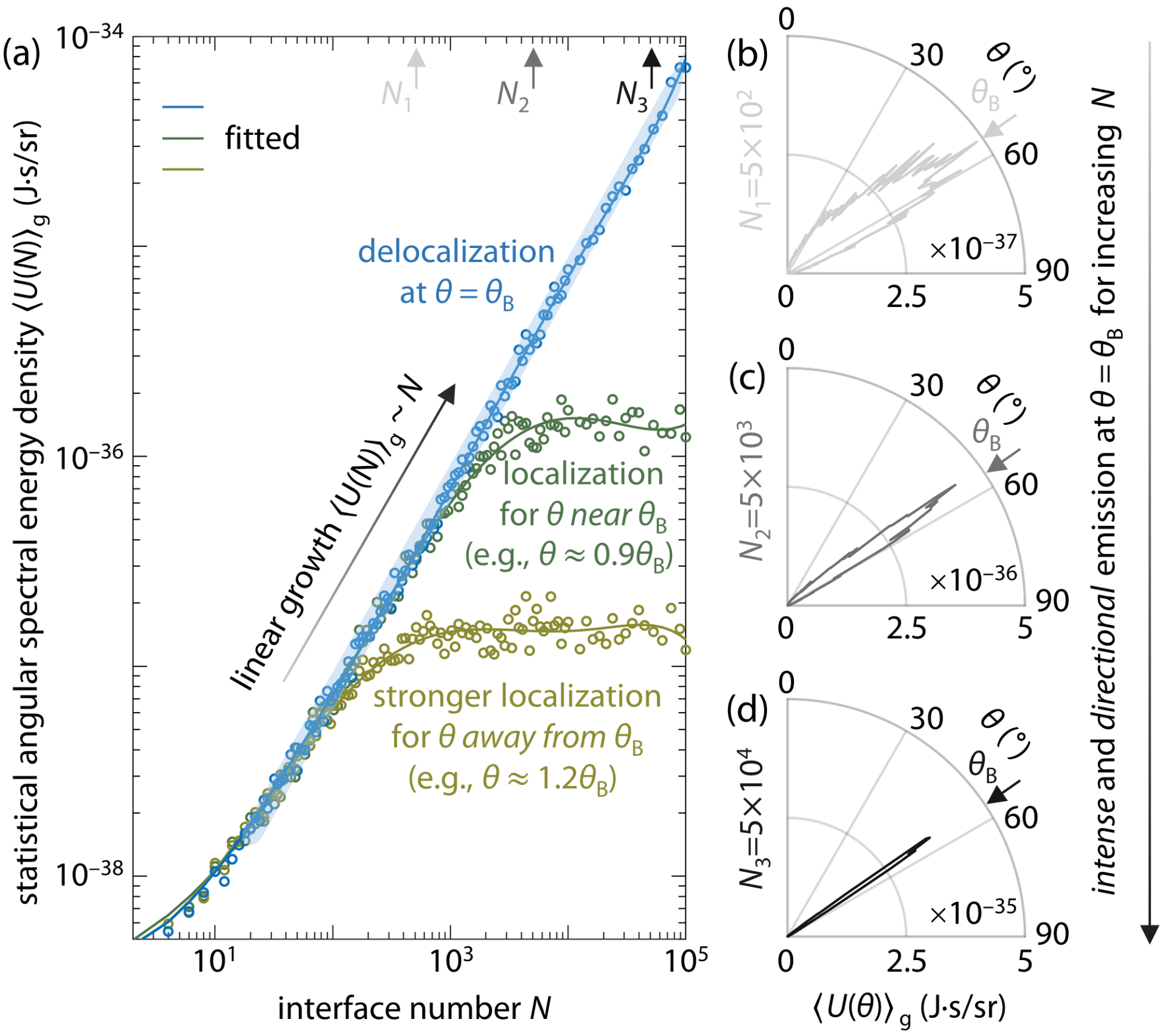

FIG. 2. Intense and directional free-electron radiation from unengineered random media via Brewster-anomaly delocalization. (a) Dependence of the configurationally averaged angular radiation energy density $\langle U(N)\rangle_{\mathrm{g}}$ on the interface number $N$ for various radiation angles. (b)–(d) Angular distribution $\langle U(\theta)\rangle_{\mathrm{g}}$ for various interface numbers. $\langle U\rangle_{\mathrm{g}} = \sqrt[M]{\prod_{i=1}^{M} U_{\mathrm{single},i}} = e^{\sum_{i=1}^{M}\ln(U_{\mathrm{single},i})/M}$ is the geometric average over $M$ configurations of the random medium, where $U_{\mathrm{single},i}$ denotes the $i$-th single configuration. For a large $N$ in (a), the radiation intensity at the non-Brewster angles saturates with increasing interfaces, while it grows linearly at the Brewster angle. This contrasting behavior leads not only to improved intensity but also to directional emission at the Brewster angle in (b)–(d), where $N$ is 500, 5000, and 50000, respectively. For illustration here and below, we use $M = 100$, the Brewster angle $\theta_{\mathrm{B}} = 55.4°$, and selected angles near and far from the Brewster angle are $49.4° \approx 0.9\theta_{\mathrm{B}}$ and $67.2° \approx 1.2\theta_{\mathrm{B}}$.

To provide deeper insight, we highlight in Fig. 3 that disorder can actually constitute a key factor in improving the directionality of free-electron radiation in random media. To begin with, Figure 3(a) shows that free-electron radiation is statistically robust against various random-medium configurations, as evidenced by globally consistent radiation features. Averaging over multiple configurations of random media suppresses speckle noise [57,84,86,87], isolating the robust signatures in Fig. 3(b). Closer inspection of the spectra [Figs. 3(a) and 3(b)] reveals two distinct bright stripes. The first appears at low disorder strength $\sigma/\lambda_0 \ll 1$ and vanishes as $\sigma$ increases, indicating the localization-mediated suppression of free-electron radiation. By contrast, the second bright stripe persists near the Brewster angle even under strong disorder $\sigma/\lambda_0 \gg 1$, confirming the delocalization of free-electron radiation in this regime.

On this basis, we extract from Fig. 3(b) the configurationally averaged radiation intensity $\langle U \rangle_{\mathrm{g}}$ as a function of disorder strength for three representative cases: away from the Brewster angle $\theta_{\mathrm{B}}$ (e.g., $\theta \approx 1.2\theta_{\mathrm{B}}$), near $\theta_{\mathrm{B}}$ (e.g., $\theta \approx 0.9\theta_{\mathrm{B}}$), and exactly at $\theta_{\mathrm{B}}$ (i.e., $\theta = \theta_{\mathrm{B}}$). Far from the Brewster angle, the radiation intensity decays rapidly with increasing disorder before eventually saturating. A similar crossover from decay to saturation arises near the Brewster angle, though the decay rate is slower. Exactly at $\theta_{\mathrm{B}}$, the intensity remains nearly constant across the entire range of disorder. Viewed differently, the disparity between the Brewster and non-Brewster regimes at low disorder and their eventual convergence at high disorder mark the evolution of free-electron radiation from its onset to its steady state. As shown in Figs. 3(d)–3(g), increasing disorder strength enhances the directionality (characterized by the angular width $\Delta\theta$), which eventually remains relatively stable in the strong-disorder regime.

Before closing, we find that the delocalized free-electron radiation mediated by Brewster anomalies exhibits a certain tolerance to variations in material permittivity in Fig. S4 and material loss in Fig. S5. On this basis, we show in Fig. S6 that realistic low-loss dispersive dielectrics are widely available for constructing layered nanostructures that support Brewster-anomaly-mediated free-electron radiation.

The constant-velocity approximation is justified because changes in the electron velocity caused by photon emission are negligible and unwanted inelastic electron scattering is effectively suppressed [see also Fig. S7].

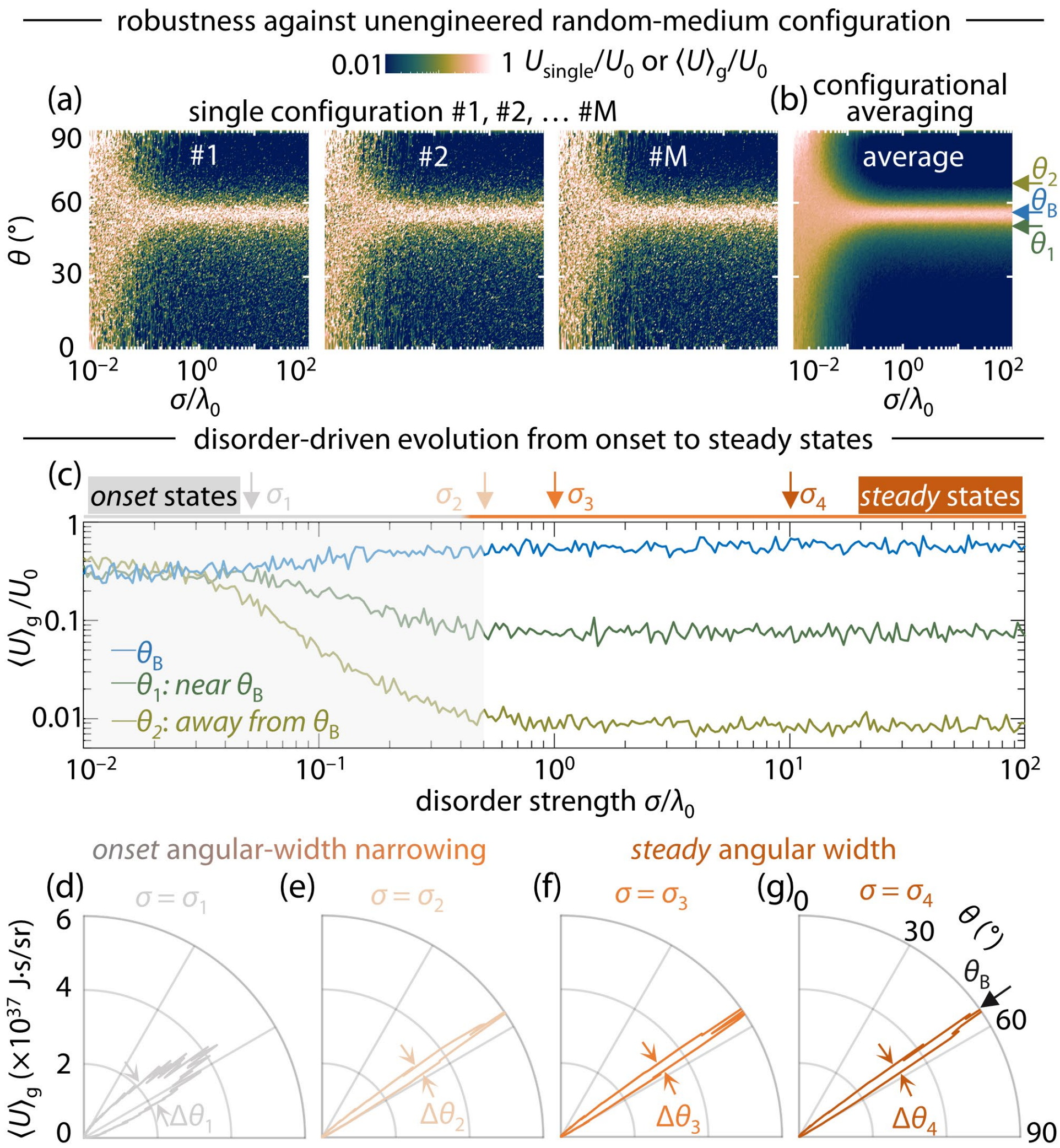

FIG. 3. Disorder-driven evolution of statistically robust free-electron radiation via Brewster-anomaly delocalization. (a),(b) Single-configuration and configurationally averaged angular spectral energy density mapped in the disorder strength $\sigma$ – radiation angle $\theta$ parameter space. The auxiliary normalization factor $U_0 = U_0(\theta)$ is a direct sum of radiation at sequential interfaces; it neglects multiple-scattering and interference effects and is thus independent of $\sigma$. (c) Dependence of $\langle U(\sigma)\rangle_g / U_0$ on the disorder strength $\sigma$ for various emission angles $\theta$. (d)–(g) Angular distribution $\langle U(\theta)\rangle_g$ for various disorder strengths. Locally varied but globally similar free-electron radiation features shown in (a) and (b) indicate its statistical robustness to the random-medium configuration. As the disorder strength increases in (c)–(g), the disorder drives an evolution of free-electron radiation from onset to steady states,

respectively characterized by narrowing and stabilizing angular widths (full width at half maximum $\Delta\theta$ of $\langle \boldsymbol{U}(\theta)\rangle_{\mathrm{g}}$).

To conclude, we have proposed a delocalization-mediated mechanism for free-electron radiation in strongly disordered, unengineered random media. This mechanism could enable the emergence of intense, directional free-electron radiation that is robust to variations in unengineered disorder, light frequency and particle velocity. Our findings overturn the belief that free-electron radiation in strongly disordered, unengineered systems is inherently random and unpredictable, establishing a viable route for controlling free-electron emission in complex photonic environments. Beyond the disorder-immune nature of Brewster-anomaly delocalization for free-electron radiation, its robustness to light frequency and electron velocity [see End Matter, Fig. 4] carries far-reaching implications. First, the achromatic and broadband light emission in random media is highly relevant to laser-based practical applications [88,89] and ultrafast optics [90,91]. Second, the robust response to electron velocity suggests a high degree of tolerance to the noise within the excitation source. The high tolerance paves the way for random radiator-based particle detectors with high signal-to-noise ratios; it also indicates the feasibility of exploring the interplay between Brewster-anomaly delocalization and more complex forms of excitations, including moving or stationary dipoles [92–94], multipoles [95], propagating pulses [31], and realistic electron beam with energy and momentum spread.

*Acknowledgements*—X.L. acknowledges support partly from National Key Research and Development Program of China under Grant No. 2025YFF0514901, the National Natural Science Foundation of China (NSFC) under Grants Nos. W2543013 and 62475227, and Zhejiang Provincial Natural Science Foundation of China under Grant No. LR26F050002. S.R. acknowledges support by the Austrian Science Fund (FWF) through project META-INCOME [10.55776/PIN7240924]. O.D.M. acknowledges support by the Simons Collaboration on Extreme Wave Phenomena Based on Symmetries (Award No. SFI-MPS-EWP-00008530-09). H.C. acknowledges support partly from the National Natural Science Foundation of China (NSFC) under Grant Nos. U25A20520 and 62475228, and the Key Research and Development Program of the Ministry of Science and Technology under Grants Nos. 2022YFA1404704 and 2022YFA1405201. Z.G. acknowledges support from Zhejiang University through the Qiushi Flying Eagle Program and thanks Nazar Pyvovar for helpful discussions on computational techniques.

## End Matter

To place our findings in the proper context, Figure 4(a) captures the uniqueness of this work by mapping it onto a general phase diagram of reported free-electron–random media interactions. Three distinct regimes emerging in the phase diagram are illustrated by paradigmatic examples in Figs. 4(b)–4(j). The Brewster-anomaly delocalization for free-electron radiation uncovered in this work falls within the strong unengineered-disorder regime (i.e., regime ① in Fig. 4(a)), which represents the general case of random media characterized by large disorder strength $\sigma$ and arbitrary characteristic size $d_0$. In the equivalent dipolar-radiation picture, this regime is characterized by both random phases within a single realization and large phase variations under multiple realizations [Fig. 4(b)]. Remarkably, we show that in this regime free-electron radiation directionality remains robust against variations in both light frequency and particle velocity, manifesting as a persistent radiation peak at the Brewster angle [Figs. 4(c) and 4(d)]. This robustness is rooted in the fact that Brewster-anomaly delocalization for free-electron radiation does not rely on structural resonance. Meanwhile, the value of the Brewster angle is independent of particle velocity and can remain consistent over a broad spectrum for non-dispersive constituent dielectrics, which are abundant in nature [96].

The robustness also distinguishes our findings from previously reported forms of free-electron radiation that are invariant to structural randomness. In the engineered-disorder regime (i.e., regime ② in Fig. 4(a)), the structural variation is pseudo-random. Specifically, the fundamental distinction between the strong, unengineered-disorder regime and engineered-disorder regime lies not in the disorder strength, but in whether the underlying randomness is discrete or continuous. For example, in the case of the so-called Brewster randomness [53] shown by a set of discrete points spanning the full range from weak to strong disorder in Fig. 4(a), only a discrete set of layer thicknesses – integer multiples of a specific value – is permitted, which locks the phase coherence [Fig. 4(e)]. Since this fixed phase coherence is easily disrupted by the variations in light frequency and particle velocity, the resulting free-electron radiation based on engineered resonances is highly sensitive [Figs. 4(f) and 4(g)]. Finally, a more conventional weak-disorder regime (i.e., regime ③ in Fig. 4(a)) exists in weakly perturbed periodic systems, such as metamaterials or photonic crystals with short-range randomness introduced by fabrication imperfections. In such systems, the phase variation is so small [Fig. 4(h)] that free-electron radiation is still governed by the Bloch-mode dispersion and thus remains sensitive to the light frequency and electron velocity [Figs. 4(i) and 4(j)]. From an alternative perspective, the contrast between free-electron radiation in the weak-disorder regime and the strong unengineered-disorder regime suggests a disorder-immune pathway via the Brewster effect to overcome the inherent chromaticity (i.e., frequency dependence) of light emission from photonic crystals.

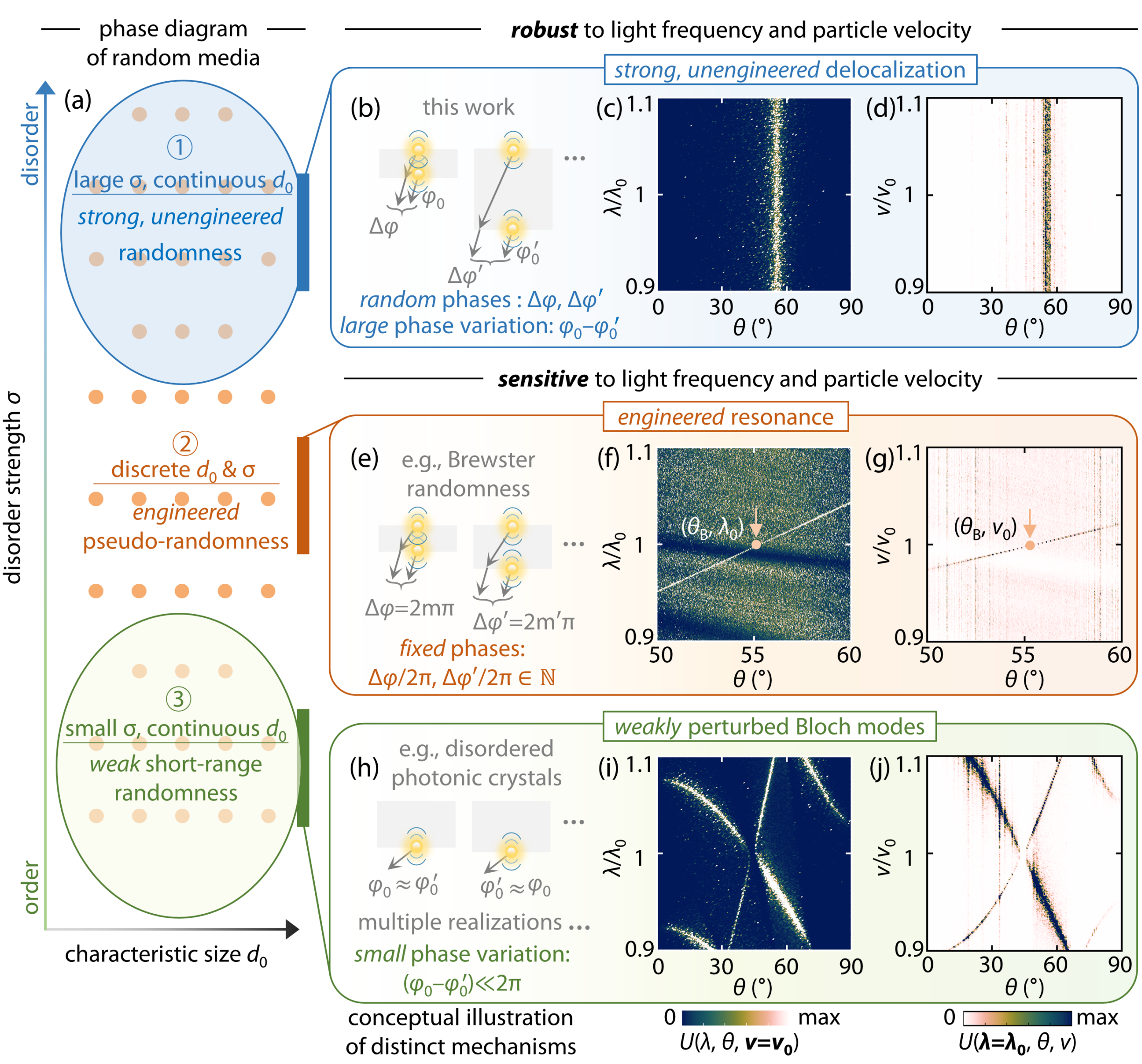

FIG. 4. Phase diagram of free-electron–random media interactions and their sensitivity to light frequency and particle velocity. (a) Phase diagram in the parameter space defined by characteristic size $d_0$ and disorder strength $\sigma$. (b)–(j) Underlying mechanisms and single-configuration angular spectral energy densities. By comparing two random configurations of a specific layer in (b), (e) and (h), distinct mechanisms are qualitatively distinguished by the phase differences between adjacent dipoles ($\Delta\varphi$ and $\Delta\varphi'$) within a single realization, and the variation in the initial dipole phases ($\varphi_0$ and $\varphi_0'$) under multiple realizations. Intriguingly, only the strong unengineered-disorder regime ① revealed in this work is robust to the variations in light frequency and particle velocity, shown by the vertical resonant lines in (c) and (d). The other conventional regimes include engineered-disorder regime ② in (f) and (g) and weak-disorder regime ③ in (i) and (j), where the free-electron radiations are featured by engineered resonances and weakly perturbed Bloch modes, respectively. For illustration, the Brewster-anomaly delocalization corresponds to strong unengineered-disorder regime in (b)–(d), where $N = 50000$. The so-called Brewster randomness [53] corresponds to engineered-disorder regime in (e)–(g), where $N = 5000$ and each layer thickness is a random odd multiple (e.g., discretely and uniformly distributed among $1,3,5,\cdots,19$) of a basic value. Disordered photonic crystals [45] correspond to weak-disorder regime in (h)–(j), where $N = 50000$ and each layer thickness follows a continuous uniform distribution over $[0.98\lambda_0, 1.02\lambda_0]$. All other structural setups in (b)–(j) are the same as those in Fig. 2.

---

Supplementary information for

# Brewster-anomaly delocalization for free-electron radiation

Zheng Gong, Xiangfeng Xi, Hongsheng Chen, Owen D. Miller, Stefan Rotter, and Xiao Lin

## Guide To the Supplementary Sections

## Derivation of free-electron radiation from layered random nanostructures

In this section, we analytically derive the free-electron radiation from a layered random nanostructure [Fig. 1(a) in the main text] within the framework of classical electrodynamics. Specifically, we consider the radiation emitted by a charge $q$ normal incident on a layered random nanostructure with $N$ interfaces dividing the space into $N+1$ regions. Without loss of generality, we assume the top (region $j=1$) and bottom ($j=N+1$) regions are vacuum, while the constituent medium in region $j$ ($j=2,3\dots N$) has a relative permittivity $\varepsilon_{\mathrm{r},j}$ and a thickness $d_j$. In addition, we shall consider uniform motion of the charge velocity $\bar{v}=\hat{z}v$, where the total energy of emitted photons is generally a very small fraction of the electron's kinetic energy [S1,S2].

### S1.1 Plane-wave expansion and wave equation

In this subsection, we introduce the plane-wave expansion as the key tool for analytically solving the of free-electron radiation from layered random nanostructures. That is,

$$X(\bar{r},t)=\int d\omega\int d\bar{k}_\perp X_{\omega,k_\perp}e^{i\bar{k}_\perp\cdot\bar{r}_\perp-i\omega t} \tag{S1}$$

where $X$ can be any component of the field or source. The wave equation in the $\omega-k_\perp$ space of a medium with permittivity $\varepsilon$ and permeability $\mu$ is [S3]

$$\left(\frac{\partial^2}{\partial z^2}+\omega^2\mu\varepsilon+\nabla_\perp^2\right)\begin{pmatrix}E_z\\H_z\end{pmatrix}=\begin{pmatrix}\left(\frac{\partial^2}{\partial z^2}+\omega^2\mu\varepsilon\right)\frac{J_z}{i\omega\varepsilon}\nabla_\perp\cdot\frac{\partial\bar{J}_\perp}{\partial z}\\-\frac{\nabla_\perp\cdot\frac{\partial}{\partial z}\bar{J}_\perp}{i\omega\mu}\end{pmatrix} \tag{S2}$$

where the subscript $\perp$ indicates the component perpendicular to the electron velocity. This notation applies for all physical quantities in the following. For example, $\bar{r}_\perp=\hat{x}x+\hat{y}y$, $\bar{k}_\perp=\hat{x}k_x+\hat{y}k_y$, and $\nabla_\perp=\nabla-\hat{z}\frac{\partial}{\partial z}$. Based on the plane-wave expansion, one has the particle current density in the $\omega-k_\perp$ space by performing the inverse transform

$$\bar{J}_{\omega,k_\perp}=\int dt\int d\bar{r}\;\bar{J}(\bar{r},t)e^{-i\bar{k}_\perp\cdot\bar{r}_\perp+i\omega t} \tag{S3}$$

where the current density in real space is

$$\bar{J}(\bar{r},t)=\hat{z}qv\delta(x)\delta(y)\delta(z-vt) \tag{S4}$$

By combing Eqs. (S2,S3), one has $\bar{J}_{\omega,\bar{k}_\perp}=\hat{z}\frac{q}{(2\pi)^3}e^{i\frac{\omega}{v}z}$. In the following, we neglect the subscripts $\omega$ and $\bar{k}_\perp$ for all quantities in the $\omega-k_\perp$ for simplicity.

It is easy to observe that in normal-incidence case here, the fields are purely transverse-magnetic (i.e. $p$-polarized) waves. On this basis, the field components perpendicular to the particle trajectory, namely $\bar{H}_\perp$ and $\bar{E}_\perp$, can be expressed via $E_z$ and $H_z$, namely

$$\bar{E}_\perp = \frac{i\omega\mu\left(\nabla_\perp \times \bar{H}_z - \bar{J}_\perp\right) + \nabla_\perp \frac{\partial E_z}{\partial z}}{\frac{\partial^2}{\partial z^2} + \omega^2 \mu\varepsilon}$$
$$\bar{H}_\perp = \frac{-i\omega\varepsilon\nabla_\perp \times \bar{E}_z - \hat{z} \times \frac{\partial \bar{J}_\perp}{\partial z} + \nabla_\perp \frac{\partial H_z}{\partial z}}{\frac{\partial^2}{\partial z^2} + \omega^2 \mu\varepsilon} \tag{S5}$$

## S1.2 Scattered field from free-electron incidence on an interface

In this subsection, we re-occur here the conventional formulation of free-electron radiation at a single interface initialized by Ginzburg and Frank and known as transition radiation [S4,S5]. We add more physical insights here by interpreting it as a scattering problem. The total electric filed is decomposed to the incident part (i.e., the charge field) and the scattered part (i.e., the radiation field), namely

$$\bar{E}^{\text{total}}\left(\bar{r}, t\right) = \bar{E}^{\text{inc}}\left(\bar{r}, t\right) + \bar{E}^{\text{sca}}\left(\bar{r}, t\right) \tag{S6}$$

Without loss of generality, we consider a single interface located at $z = z_j$ separating regions $j$ and $j + 1$. The *incident* charge field $E_j^{\text{inc}}$ at region $j$ is the non-homogeneous solution of Eq. (S2), where the operators $\nabla_\perp^2$ and $\frac{\partial^2}{\partial z^2}$ are replaced by $-k_\perp^2$ and $-\left(\frac{\omega}{v}\right)^2$, respectively. Then one has

$$E_{z,j}^{\text{inc}} = \frac{-iq}{\omega\varepsilon_0 (2\pi)^3} \frac{1 - \frac{c^2}{v^2 \varepsilon_{\text{r},j}}}{\varepsilon_{\text{r}.j} - \frac{c^2}{v^2} - \frac{k_\perp^2 c^2}{\omega^2}} e^{i\frac{\omega}{v}z} \tag{S7}$$

By plane-wave expansion, one has real space-time incident charge field in $(\rho, \phi, z)$ cylindrical coordinates:

$$E_{z,j}^{\text{inc}}(r, t) = \int_{-\infty}^{\infty} d\omega \frac{-q}{8\pi\omega\varepsilon_0\varepsilon_{\text{r},j}} \left(\frac{\omega^2}{c^2}\varepsilon_{\text{r},j} - \frac{\omega^2}{v^2}\right) H_0^{(1)}\left(\sqrt{\frac{\omega^2}{c^2}\varepsilon_{\text{r},j} - \frac{\omega^2}{v^2}}\,\rho\right) e^{i\frac{\omega}{v}z - i\omega t} \tag{S8}$$

where $c$ is the speed of light in free space, $\varepsilon_0$ is the free-space permittivity, $H_n^{(1)}$ represents the first kind of Hankel function of $n$ order.

The *scattered* radiation fields from the interface between regions $j$ and $j + 1$ comes from the homogeneous term of Eq. (S4). We write them with undetermined radiation factors $a_{j|j+1}^-$ ($a_{j|j+1}^+$) indicating backward (forward) scattering into regions $j$ ($j + 1$), namely

$$E_{z,j}^{\mathrm{sca}} = \frac{iq}{\omega\varepsilon_0(2\pi)^3} a_{j|j+1}^{-} e^{-ik_{z,j}(z-d_j)} \tag{S9}$$

$$E_{z,j+1}^{\mathrm{sca}} = \frac{iq}{\omega\varepsilon_0(2\pi)^3} a_{j|j+1}^{+} e^{ik_{z,j+1}(z-d_j)} \tag{S10}$$

Similarly, by plane-wave expansion, the corresponding real-space fields are written as

$$E_{z,j}^{\mathrm{sca}}(\bar{r},t) = \int_{-\infty}^{\infty} d\omega \int_{0}^{\infty} dk_\perp \frac{iq}{\omega\varepsilon_0(2\pi)^3} a_{j|j+1}^{-} \left(2\pi k_\perp J_0(k_\perp\rho)\right) e^{-ik_{z,j}(z-z_j)-i\omega t} \tag{S11}$$

$$E_{z,j+1}^{\mathrm{sca}}(\bar{r},t) = \int_{-\infty}^{\infty} d\omega \int_{0}^{\infty} dk_\perp \frac{iq}{\omega\varepsilon_0(2\pi)^3} a_{j|j+1}^{+} \left(2\pi k_\perp J_0(k_\perp\rho)\right) e^{ik_{z,j+1}(z-z_j)-i\omega t} \tag{S12}$$

where the component of wavevector along $z$ direction in region $j$ is $k_{z,j} = \sqrt{\frac{\omega^2}{c^2}\varepsilon_{\mathrm{r},j} - k_\perp^2}$.

At this point, Eqs. (S8,S11,S12) give the main results for the scattered field from free-electron incidence on a single interface (i.e., transition radiation), except the undetermined radiation factors $a_{j|j+1}^{-}$ and $a_{j|j+1}^{+}$. To this end, by matching the boundary condition at interface $z = z_j$, one has

$$\begin{aligned} a_{j|j+1}^{-} = a_{j|j+1}^{-,0} e^{\frac{i\omega}{v}d_j} &= \frac{\frac{v}{c}\cdot\frac{k_\perp^2 c^2}{\omega^2\varepsilon_{\mathrm{r},j}}\left(\varepsilon_{\mathrm{r},j+1}-\varepsilon_{\mathrm{r},j}\right)\left(1-\frac{v^2}{c^2}\varepsilon_{\mathrm{r},j}+\frac{v}{c}\cdot\frac{k_{z,j+1}}{\omega/c}\right)}{\left(1-\frac{v^2}{c^2}\varepsilon_{\mathrm{r},j}+\frac{k_\perp^2 v^2}{\omega^2}\right)\left(1+\frac{v}{c}\cdot\frac{k_{z,j+1}}{\omega/c}\right)\left(\varepsilon_{\mathrm{r},j}\frac{k_{z,j+1}}{\omega/c}+\varepsilon_{\mathrm{r},j+1}\frac{k_{z,j}}{\omega/c}\right)} e^{i\frac{\omega}{v}z_j} \\ a_{j|j+1}^{+} = a_{j|j+1}^{+,0} e^{\frac{i\omega}{v}d_j} &= \frac{\frac{v}{c}\cdot\frac{k_\perp^2 c^2}{\omega^2\varepsilon_{\mathrm{r},j+1}}\left(\varepsilon_{\mathrm{r},j+1}-\varepsilon_{\mathrm{r},j}\right)\left(1-\frac{v^2}{c^2}\varepsilon_{\mathrm{r},j+1}-\frac{v}{c}\cdot\frac{k_{z,j}}{\omega/c}\right)}{\left(1-\frac{v^2}{c^2}\varepsilon_{\mathrm{r},j+1}+\frac{k_\perp^2 v^2}{\omega^2}\right)\left(1-\frac{v}{c}\cdot\frac{k_{z,j}}{\omega/c}\right)\left(\varepsilon_{\mathrm{r},j}\frac{k_{z,j+1}}{\omega/c}+\varepsilon_{\mathrm{r},j+1}\frac{k_{z,j}}{\omega/c}\right)} e^{i\frac{\omega}{v}z_j} \end{aligned} \tag{S13}$$

where $a_{j|j+1}^{-,0}$ ($a_{j|j+1}^{+,0}$) is the auxiliary quantity representing backward (forward) radiation factors at the interface located at $z = 0$.

## S1.3 Free-electron radiation from layered random nanostructures

In this subsection, we derive the field and energy spectrum of free-electron radiation from layered random nanostructures. Two basic steps are needed for this purpose: first, we shall account for the multiple scattering of the initial emission at the interface; second, we obtain the total field by superposing the contribution from all interfaces using the transfer-matrix formalism.

### S1.3.1 Multiple scattering

In this subsubsection, we show how the multiple scattering in layered random system affects the transition-radiation filed distribution. Specifically, the contribution of the forward (backward) transition radiation at the $j-1|j$ ($j|j+1$) interface to the field in region $j$ is respectively given by

$$\mathrm{TR}_j^+ \rightarrow \text{region } j = A_{j,0}^+ \cdot \left[e^{ik_{z,j}(z-z_{j-1})} + \tilde{R}_{j|j+1} e^{2ik_{z,j}(z_j-z_{j-1})} e^{-ik_{z,j}(z-z_{j-1})}\right] \tag{S14}$$

$$\mathrm{TR}_j^- \rightarrow \text{region } j = A_{j,o}^- \cdot \left[e^{-ik_{z,j}(z-z_j)} + \tilde{R}_{j|j-1} e^{+ik_{z,j}(z-z_j)} \cdot e^{-2ik_{z,j}(z_{j-1}-z_j)}\right] \tag{S15}$$

where $A_{j,o}^+$ and $A_{j,0}^-$ are the modified transition-radiation factor in region $j$ from the adjacent interfaces located at $z=z_{j-1}$ and $z=z_j$, after accounting for the multiple scattering effect. Fundamentally, $M_j$ characterizes how the radiation factors of the secondary transition-radiation source at the $j-1|j$ and $j|j+1$ interfaces are modified by multiple scattering [S6], namely

$$A_{j,0}^+ = \frac{iq}{\omega\varepsilon_0(2\pi)^3} \cdot a_{j-1|j}^+ M_j\ , \quad A_{j,o}^- = \frac{iq}{\omega\varepsilon_0(2\pi)^3} \cdot a_{j|j+1}^- M_j \tag{S16}$$

$$M_j = \frac{1}{1 - \tilde{R}_{j|j+1}\tilde{R}_{j|j-1} e^{2ik_{z,j}(z_j-z_{j-1})}}\ , \quad j = 2, 3, \ldots N \tag{S17}$$

$\tilde{R}_{j|j-1}$ and $\tilde{R}_{j|j+1}$ are the generalized reflection coefficients for $p$-polarized waves at the $j-1|j$ interface (from region $j-1$ to region $j$) and the $j|j+1$ interface (from region $j$ to region $j+1$), respectively. The generalized reflection coefficients could be obtained by using an iteration relationship [S6], namely

$$\begin{aligned} \tilde{R}_{j|j+1} &= R_{j|j+1} + \frac{T_{j|j+1}\tilde{R}_{j+1|j+2}T_{j+1|j} e^{2ik_{z,j+1}(z_{j+1}-z_j)}}{1 - R_{j+1|j}\tilde{R}_{j+1|j+2} e^{2ik_{z,j+1}(z_{j+1}-z_j)}} \\ \tilde{R}_{j|j-1} &= R_{j|j-1} + \frac{T_{j|j-1}\tilde{R}_{j-1|j-2}T_{j-1|j} e^{2ik_{z,j-1}(z_{j-1}-z_{j-2})}}{1 - R_{j-1|j}\tilde{R}_{j-1|j-2} e^{2ik_{z,j-1}(z_{j-1}-z_{j-2})}} \end{aligned} \tag{S18}$$

The iteration in Eq. (S18) can start from the first or the last interface due to the facts that $\tilde{R}_{N|N+1} = R_{N|N+1}$ and $\tilde{R}_{2|1} = R_{2|1}$. The transmission and reflection coefficients for the field $E_z$ at interface $z = d_j$ from region $j$ to region $j+1$ can be expressed as $T_{j\,|\,j+1} = T^{\mathrm{TM}} \frac{\varepsilon_{\mathrm{r},j}}{\varepsilon_{\mathrm{r},j+1}} = \frac{2\varepsilon_{\mathrm{r},j+1}k_{z,j}}{\varepsilon_{\mathrm{r},j+1}k_{z,j}+\varepsilon_{\mathrm{r},j}k_{z,j+1}} \cdot \frac{\varepsilon_{\mathrm{r},j}}{\varepsilon_{\mathrm{r},j+1}}$ and $R_{j\,|\,j+1} = R^{\mathrm{TM}} = \frac{\varepsilon_{\mathrm{r},j+1}k_{z,j}-\varepsilon_{\mathrm{r},j}k_{z,j+1}}{\varepsilon_{\mathrm{r},j+1}k_{z,j}+\varepsilon_{\mathrm{r},j}k_{z,j+1}}$, where $R^{\mathrm{TM}}, T^{\mathrm{TM}}$ are the generally used reflection and transmission coefficients of TM waves defined for the magnetic field $H_\perp$. Similarly, from region $j+1$ to region $j$, we have $T_{j+1\,|\,j} = \frac{2\varepsilon_{\mathrm{r},j}k_{z,j+1}}{\varepsilon_{\mathrm{r},j+1}k_{z,j}+\varepsilon_{\mathrm{r},j}k_{z,j+1}} \cdot \frac{\varepsilon_{\mathrm{r},j+1}}{\varepsilon_{\mathrm{r},j}}$ and $R_{j+1\,|\,j} = -R_{j\,|\,j+1} = -\frac{\varepsilon_{\mathrm{r},j+1}k_{z,j}-\varepsilon_{\mathrm{r},j}k_{z,j+1}}{\varepsilon_{\mathrm{r},j+1}k_{z,j}+\varepsilon_{\mathrm{r},j}k_{z,j+1}}$.

#### S1.3.2 Superposition from all secondary sources

In this subsubsection, we formulate the total free-electron radiation as the superposition from all secondary sources (i.e., forward and backward transition radiations at each interface). The total scattered field in region $j$ originates from the interfaces above and below this region and comprises the following four distinct parts.

We first consider the Parts 1-2 contributed by the interfaces above region $j$.

- **Part 1: The forward transition radiation at the $m-1|m$ ($2 \le m \le j$) interface**

In light of Eq. (S14), we write this part as

$$\mathrm{TR}_m^+ \rightarrow \text{region } j = C_{m|j}^+ \cdot \left[e^{ik_{z,j}(z-z_{j-1})} + \tilde{R}_{j|j+1} e^{2ik_{z,j}(z_j - z_{j-1})} e^{-ik_{z,j}(z-z_{j-1})}\right] \tag{S19}$$

Equation (S14) gives the trivial case of $m = j$ with $C_{m|j}^+ = C_{j|j}^+ = A_{j,0}^+$. The general solution of $C_{m|j}^+$ ($2 \le m \le j-1$) is obtained using a generalized transmission coefficient from region $m$ to region $j$ $\tilde{T}_{m|j}^{j>m}$ as follows. Specifically, at the $z = z_m$ interface, we have

$$C_{m|m+1}^+ e^{ik_{z,m+1}(z_m - z_{m-1})} = A_{m,0}^+ e^{ik_{z,m}(z_m - z_{m-1})} \cdot S_{m|m+1} \tag{S20}$$

$$S_{m|m+1} = \frac{T_{m|m+1}}{1 - R_{m+1|m} \tilde{R}_{m+1|m+2} e^{i2k_{z,m+1}(z_{m+1} - z_m)}} \tag{S21}$$

Equations (S20,S21) indicate that the value of $C_{m|m+1}^+$ is determined by $A_{m,0}^+$ and the transmission at the $m|m+1$ interface. Recursively matching the boundary conditions from region $m$ to region $j$, we have

$$\prod_{n=m}^{n=j-1} C_{m|n+1}^+ e^{ik_{z,n+1}(z_n - z_{m-1})} = \prod_{n=m}^{n=j-1} C_{m|n}^+ e^{ik_{z,n}(z_n - z_{m-1})} \cdot S_{n|n+1} \tag{S22}$$

By further using $C_{m|m}^+ = A_{m,0}^+$, one has

$$C_{m|j}^+ e^{ik_{z,j}(z_{j-1} - z_{m-1})} = \tilde{T}_{m|j}^{j>m} \cdot A_{m,0}^+ \tag{S23}$$

$$\tilde{T}_{m|j}^{j>m} = \prod_{n=m}^{n=j-1} S_{n|n+1}\, e^{ik_{z,n}(z_n - z_{n-1})} \tag{S24}$$

where $\tilde{T}_{m|j}^{j>m}$ is the generalized transmission coefficient from region $m$ to region $j$.

- **Part 2: The backward transition radiation at the $m|m+1$ ($2 \le m \le j$) interface**

The part 2 is defined by

$$\mathrm{TR}_m^- \rightarrow \text{Region } j = D_{m|j}^+ \left[e^{ik_{z,j}(z-z_m)} + \tilde{R}_{j|j+1} e^{2ik_{z,j}(z_j - z_m)} e^{-ik_{z,j}(z-z_m)}\right] \tag{S25}$$

$$D_{m|j}^+ e^{ik_{z,j}(z_{j-1} - z_m)} = \tilde{T}_{m|j}^{j>m} BR_{m,o}^+ \cdot e^{ik_{z,m}(z_{m-1} - z_m)} \tag{S26}$$

where the calculation procedure for the unknown factor $D^{+}_{m|j}$ is exactly the same as that of $C^{+}_{m|j}$ except the quantity $BR^{+}_{m,o}$ (the counterpart of $A^{+}_{m,0}$ in part 1) now originating the *backward* transition radiation at the $m|m+1$ interface. To solve $BR^{+}_{m,o}$, we transform the emitted field in region $m$ from the backward transition radiation at the $m|m+1$ interface [S6] as

$$\begin{aligned}\mathrm{TR}^{-}_{m}\rightarrow \text{Region } m &= A^{-}_{m,o}\left[e^{-ik_{z,m}(z-z_m)}+\tilde{R}_{m|m-1}e^{ik_{z,m}(z-z_m)}\cdot e^{-2ik_{z,m}(z_{m-1}-z_m)}\right]\\ &= B^{-}_{m,o}e^{-ik_{z,m}(z-z_m)}+BR^{+}_{m,o}\left[e^{+ik_{z,m}(z-z_m)}+\tilde{R}_{m|m+1}e^{-ik_{z,m}(z-z_m)}e^{+2ik_{z,m}(z_m-z_m)}\right]\end{aligned} \tag{S27}$$

By substituting the relation of $M_m = 1 + M_m\tilde{R}_{m|m+1}\tilde{R}_{m|m-1}e^{2ik_{z,m}(z_m-z_{m-1})}$, we obtain that

$$BR^{+}_{m,o} = A^{-}_{m,o}\tilde{R}_{m|m-1}e^{-2ik_{z,m}(z_{m-1}-z_m)} \tag{S28}$$

In the following, we consider the Parts 3-4 contributed by the interfaces below region $j$.

➢ **Part 3: Backward transition radiation at the $m|m+1$ ($j\le m\le N$) interface**

By generalizing Eq. (S15), we write this part as

$$TR^{-}_{m}\rightarrow region\ j = D^{-}_{m|j}\left[e^{-ik_{z,j}(z-z_m)}+\tilde{R}_{j|j-1}e^{+ik_{z,j}(z-z_m)}e^{-2ik_{z,j}(z_{j-1}-z_m)}\right] \tag{S29}$$

where $D^{-}_{m|m} = A^{-}_{m,0}$; and for $j < m \le N$, $D^{-}_{m|j}$ is obtained by performing the exactly the similar procedure to part 1, one has

$$D^{-}_{m|j}e^{-ik_{z,j}(z_j-z_m)} = \tilde{T}^{j<m}_{m|j}\cdot A^{-}_{m,o} \tag{S30}$$

$$\tilde{T}^{j<m}_{m|j} = \prod_{n=j}^{n=m-1} S_{n+1|n}\, e^{-ik_{z,n+1}(z_n-z_{n+1})} \tag{S31}$$

$$S_{m|m-1} = \frac{T_{m|m-1}}{1-R_{m-1|m}\tilde{R}_{m-1|m-2}e^{2ik_{z,m-1}(z_{m-1}-z_{m-2})}} \tag{S32}$$

➢ **Part 4: Forward transition radiation at the $m-1|m$ ($j+1\le m\le N$) interface**

The part 4 is defined by

$$TR^{+}_{m}\rightarrow region\ j = C^{-}_{m|j}\left[e^{-ik_{z,j}(z-z_{m-1})}+\tilde{R}_{j|j-1}e^{+ik_{z,j}(z-z_{m-1})}e^{-2ik_{z,j}(z_{j-1}-z_{m-1})}\right] \tag{S33}$$

$$C^{-}_{m|j}e^{-ik_{z,j}(z_j-z_{m-1})} = \tilde{T}^{j<m}_{m|j}\cdot AR^{-}_{m,o}e^{-ik_{z,m}(z_m-z_{m-1})} \tag{S34}$$

where $AR^{-}_{m,o}$ is obtained by following the similar transformation to $BR^{+}_{m,o}$, namely

$$\begin{aligned}\mathrm{TR}^{+}_{m}\rightarrow \text{Region } m &= B^{+}_{m,o}e^{ik_{z,m}(z-z_{m-1})}+AR^{-}_{m,o}\left[e^{-ik_{z,m}(z-z_{m-1})}+\tilde{R}_{m|m-1}e^{ik_{z,m}(z-z_{m-1})}\right]\\ &= A^{+}_{m,o}M_m\tilde{R}_{m|m+1}e^{2ik_{z,m}(z_m-z_{m-1})}\end{aligned} \tag{S35}$$

➢ **Total field distribution from the superposition of all four parts**

The total electric field in region $j$ ($1\le j\le N+1$) can now be expressed as

$$E_{z,j}^{\rm sca} = \begin{cases} \frac{iq}{\omega\varepsilon_0(2\pi)^3} \cdot a_1 \cdot e^{-ik_{z,1}(z-z_1)} & (j=1) \\ \frac{iq}{\omega\varepsilon_0(2\pi)^3} \left[a_j^- \cdot e^{-ik_{z,j}(z-z_j)} + a_j^+ \cdot e^{ik_{z,j}(z-z_{j-1})}\right] & (2 \le j \le N) \\ \frac{iq}{\omega\varepsilon_0(2\pi)^3} \cdot a_{N+1} \cdot e^{ik_{z,N+1}(z-z_N)} & (j=N+1) \end{cases} \tag{S36}$$

where $a_j^+$ ($a_j^-$) represents the generalized forward (backward) radiation factor in region $j$ and can be determined by superposing parts 1-4. The real-space distribution could be further obtained by the plane-wave expansion defined in Eq. (S1)

$$\begin{gathered} \bar{E}_z(\bar{r},t) = \bar{E}_z^{\rm inc}(\bar{r},t) + \bar{E}_z^{\rm sca}(\bar{r},t) \\ \bar{E}_{z,j}^{\rm inc}(\bar{r},t) = \hat{z}\int_{-\infty}^{\infty} d\omega \frac{-q}{8\pi\omega\varepsilon_0\varepsilon_{{\rm r},j}}\left(\frac{\omega^2}{c^2}\varepsilon_{{\rm r},j} - \frac{\omega^2}{v^2}\right) H_0^{(1)}\left(\rho\sqrt{\frac{\omega^2}{c^2}\varepsilon_{{\rm r},j} - \frac{\omega^2}{v^2}}\right) e^{i\frac{\omega}{v}z-\omega t} \\ \bar{E}_{z,j}^{\rm sca}(\bar{r},t) = \hat{z}\int_{-\infty}^{\infty} d\omega \int_0^{\infty} dk_\perp \frac{iq}{\omega\varepsilon_0(2\pi)^3}\left(2\pi k_\perp J_0(k_\perp\rho)\right)\left[a_j^- e^{-ik_{z,j}(z-d_j)-i\omega t} + a_j^+ e^{ik_{z,j}(z-d_{j-1})-i\omega t}\right] \end{gathered} \tag{S37}$$

where $a_{N+1}^- = 0$ and $a_1^+ = 0$. Compared to Eqs. (S6-S11), only the radiation factors are modified. As a simple example, the radiation in region $N+1$ originates solely from all the interfaces above region $N+1$, namely $\left(\sum_{m=2}^{j} {\rm TR}_m^+ \rightarrow \text{Region } j\right) + \left(\sum_{m=2}^{j-1} {\rm TR}_m^- \rightarrow \text{Region } j\right)$. Further substituting the corresponding radiation factors obtained, one has the radiation factor $a_{N+1}$ expressed as

$$\begin{aligned} a_{N+1} &= \frac{\omega\varepsilon_0(2\pi)^3}{iq} \cdot \left[A_{N+1,0}^+ + \sum_{m=2}^{N}\left(C_{m|N+1}^+ e^{ik_{z,N+1}(z_N - z_{m-1})} + D_{m|N+1}^+ e^{ik_{z,N+1}(z_N - z_m)}\right)\right] \\ &= \frac{\omega\varepsilon_0(2\pi)^3}{iq} \cdot \left[A_{N+1,0}^+ + \sum_{m=2}^{N}\left(\tilde{T}_{m|N+1}^{N+1>m} \cdot A_{m,0}^+ + \tilde{T}_{m|N+1}^{N+1>m} \cdot BR_{m,o}^+ e^{ik_{z,m}(z_{m-1}-z_m)}\right)\right] \end{aligned} \tag{S38}$$

Similarly, one has the backward radiation factor in region 1 $a_1$ written as

$$a_1 = \frac{\omega\varepsilon_0(2\pi)^3}{iq}\left[A_{1,o}^- + \sum_{m=2}^{N}\left(\tilde{T}_{m|1}^{1<m} \cdot A_{m,o}^- + \tilde{T}_{m|1}^{1<m} \cdot AR_{m,o}^- e^{-ik_{z,m}(z_m - z_{m-1})}\right)\right] \tag{S39}$$

### S1.3.3 Angular energy spectrum

In this subsection, we calculate the (angular) energy spectrum of free-electron radiation from layered random nanostructures. The backward angular spectral energy density $U_{\rm B}(\theta, v)$ can be expressed as

$$W_1 = \int_0^{+\infty} d\omega \int_0^{\frac{\pi}{2}} d\theta\, U_{\rm B}(\theta, v) \cdot (2\pi\sin\theta) \tag{S40}$$

where the backward radiation angle $\theta$ is made by the wavevector $\bar{k}$ and $-\bar{v}$ as illustrated in Fig. 1(a). In the top (bottom) region, it is sufficient to calculate the energy of the radiation field $E^{\rm sca}$ asymptotically as $t \rightarrow$ $\infty$, at which the radiated wave train is already at a great distance to the interface and is therefore separated from the incident charge field $E^{\rm inc}$. This way, the total energy of the scattered field is obtained by taking account the electric and magnetic energy density (which are equal) over the whole space,

$$W_1 = \varepsilon_1 \lim_{t\to\infty} \int d\bar{r} \left|\bar{E}_1^{\rm sca}(\bar{r},t)\right|^2 = \varepsilon_1 \lim_{t\to\infty} \int d\bar{r}_\perp \int_{-\infty}^{\infty} dz \left|\bar{E}_1^{\rm sca}(\bar{r},t)\right|^2 \tag{S41}$$

$$\left|\bar{E}_1^{\rm sca}(\bar{r},t)\right|^2 = \int d\bar{k}_\perp d\bar{k}'_\perp d\omega d\omega' \, \bar{E}^{\rm sca}_{\omega,\bar{k}_\perp,1}(z) \left(\bar{E}^{\rm sca}_{\omega',\bar{k}'_\perp,1}(z)\right)^* \cdot e^{i\left[\left(\bar{k}_\perp-\bar{k}'_\perp\right)\cdot\bar{r}_\perp-(\omega-\omega')t\right]} \tag{S42}$$

By substituting Eq. (S42) into (S41) and integrating over $d\bar{r}_\perp dz d\bar{k}'_\perp d\omega'$ in Eq. (S43), one has

$$W_1 = 2\int_0^{+\infty} \int \varepsilon_1 |a_1|^2 \left(\frac{q}{\omega\varepsilon_0(2\pi)^3}\right)^2 \frac{\omega^2\sqrt{\varepsilon_{\rm r,1}}}{ck_\perp^2} \sqrt{1-\frac{k_\perp^2c^2}{\omega^2\varepsilon_{\rm r,1}}}(2\pi)^3\, d\bar{k}_\perp d\omega \tag{S43}$$

For all emitted photons, the integration over $d\bar{k}_\perp$ should be taken under the condition of $k_\perp^2 \le \frac{\omega^2}{c^2}\varepsilon_{\rm r,1}$. From Eqs. (S41,S44), backward angular spectral energy density can be obtained as

$$U_{\rm B} = \frac{\varepsilon_{\rm r,1}^{3/2} q^2 \cos^2\theta\, |a_1|^2}{4\pi^3\varepsilon_0 c \sin^2\theta} \tag{S44}$$

By following the identical procedure, the forward angular spectral energy density $U_{\rm F}(\theta, v)$ is obtained as

$$U_{\rm F} = \frac{\varepsilon_{{\rm r},N+1}^{3/2} q^2 \cos^2\theta\, |a_{N+1}|^2}{4\pi^3\varepsilon_0 c \sin^2\theta} \tag{S45}$$

The radiation angle $\theta$ for the forward radiation is made by the wavevector $\bar{k}$ and the electron velocity $\bar{v}$.

Finally, the total angular spectral energy density used throughout the main text is $U = U_{\rm B} + U_{\rm F}$.

## S2 Localization length for layered random nanostructures

In this section, we analytically derive the localization length for layered random nanostructures. First, we obtain the transmission coefficient by following Born's characteristic matrix method for layered media [S7]. Second, we provide analytical and numerical results for the localization length, namely the critical system size required for the localization of light inside layered random media [S8–S11].

- **Generalized transmission coefficient**

The layered structure [see also Fig. 1(a) in the main text] consists of $N + 1$ regions separated by $N$ interfaces. Each region $j$ ($1 \le j \le N + 1$) has a permittivity $\varepsilon_j = \varepsilon_0\varepsilon_{{\rm r},j}$ and permeability $\mu_j = \mu_0\mu_{{\rm r},j}$. The thickness of region $j$ ($2 \le j \le N$) is $d_j$. Without loss of generality, consider a $p$-polarized plane wave incident from region 1 on the layered random nanostructure. The plane-wave field distribution at each region $j$ $(2 \leq j \leq N)$ is written as

$$
\begin{aligned}
\overline{H}_j &= \hat{y} \cdot \left( a_n e^{ik_{z,j}(z - z_{j-1})} + b_n e^{-ik_{z,j}(z - z_{j-1})} \right) e^{ik_x x} \\
\overline{E}_j &= \hat{x} \cdot \frac{k_{z,j}}{\omega \varepsilon_j} \left( a_n e^{ik_{z,j}(z - z_{j-1})} - b_n e^{-ik_{z,j}(z - z_{j-1})} \right) e^{ik_x x}
\end{aligned}
\tag{S46}
$$

Similarly, for the substrate and superstrate regions, one has

$$
\begin{aligned}
\overline{H}_1 &= \hat{y} \cdot \left( e^{ik_{z,1} z} + \widetilde{R}\, e^{-ik_{z,1} z} \right) e^{ik_x x} \\
\overline{E}_1 &= \hat{x} \cdot \frac{k_{z,1}}{\omega \varepsilon_1} \left( e^{ik_{z,1} z} - \widetilde{R}\, e^{-ik_{z,1} z} \right) e^{ik_x x}
\end{aligned}
\tag{S47}
$$

$$
\begin{aligned}
\overline{H}_{N+1} &= \hat{y} \cdot \widetilde{T}\, e^{ik_{z,N+1} z} e^{ik_x x} \\
\overline{E}_{N+1} &= \hat{x} \cdot \frac{k_{z,N+1}}{\omega \varepsilon_{N+1}} \widetilde{T}\, e^{ik_{z,N+1} z} e^{ik_x x}
\end{aligned}
\tag{S48}
$$

where $\widetilde{R}$ is the generalized reflection coefficient at the first interface, and $\widetilde{T}$ is the generalized transmission coefficient from region 1 to region $N+1$.

At this point, we derive the characteristic matrix $\overline{\overline{M}}_j$ for a single spatial slab extending from $z = z_{j-1}$ to $z = z_j$, which relates the field values at its two interfaces, namely

$$
\begin{bmatrix} H_{j-1}(z_{j-1}) \\ E_{j-1}(z_{j-1}) \end{bmatrix} = \overline{\overline{M}}_j \begin{bmatrix} H_{j+1}(z_j) \\ E_{j+1}(z_j) \end{bmatrix}
\tag{S49}
$$

By directly enforcing the boundary conditions for the continuity of $\overline{H}_j$ and $\overline{E}_j$ and rewriting them in a matrix form, one can easily obtain

$$
\overline{\overline{M}}_j = \begin{bmatrix} \cos(k_{z,j} d_j) & -i \dfrac{\omega \varepsilon_j}{k_{z,j}} \sin\left(k_{z,j} d_j\right) \\ -i \dfrac{k_{z,j}}{\omega \varepsilon_j} \sin\left(k_{z,j} d_j\right) \big/ \eta_j & \cos(k_{z,j} d_j) \end{bmatrix}
\tag{S50}
$$

By multiplying Eq. (S49) iteratively for all layers, we have

$$
\begin{bmatrix} H_{N+1}(z_{N+1}) \\ E_{N+1}(z_{N+1}) \end{bmatrix} = \overline{\overline{N}} \begin{bmatrix} H_1(z_1) \\ E_1(z_1) \end{bmatrix}, \quad \text{where } \overline{\overline{N}} = \prod_{j=2}^{N} \overline{\overline{M}}_j
\tag{S51}
$$

By substituting Eqs. (S47,S48) into Eq. (S51), we finally arrive at the generalized coefficient:

$$
\widetilde{T} = \frac{2}{\left( N_{11} + N_{12} \dfrac{k_{z,N+1}}{\omega \varepsilon_{N+1}} \right) + \dfrac{\omega \varepsilon_1}{k_{z,1}} \left( N_{21} + N_{22} \dfrac{k_{z,N+1}}{\omega \varepsilon_{N+1}} \right)}
\tag{S52}
$$

where $k_{z,1} = k_{z,N+1} = \frac{\omega}{c} \cos\theta$ for vacuum substrate and superstrate, and $N_{mn}$ denotes the matrix element.

## ➢ **Localization length**

The localization length is calculated numerically using its standard definition [S8–S11] as

$$\ell_{\text{loc}} = \frac{L}{-\langle \ln|\widetilde{T}|\rangle} \tag{S53}$$

where $\ell_{\text{loc}}$ is the localization length, $L = \sum_{j=2}^{N}\langle d_j\rangle$ is the average system size, and $\langle \dots \rangle$ is the arithmetic mean, representing the configurational average over multiple realizations.

We show in Fig. S1 that both localization and delocalization of light occur in layered random nanostructure considered in the main text. Specifically, as shown in Fig. S1, localization occurs when the localization length is smaller than the system size, e.g., at the non-Brewster angles. Yet, for emission angle near or exactly at the Brewster angle, the localization length could be larger than the system size, creating delocalized states.

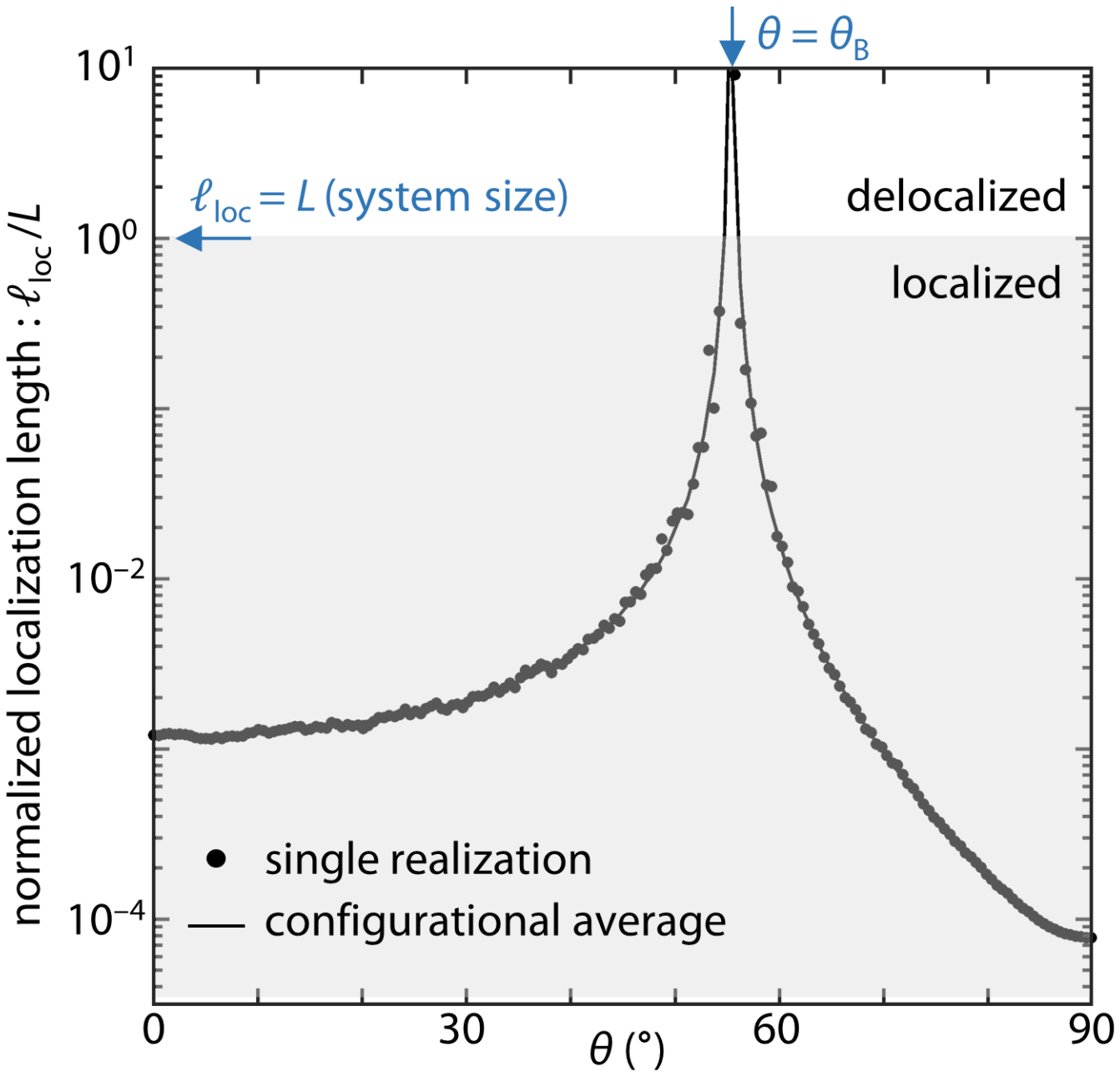

FIG. S1. Localization and Brewster-anomaly delocalization in layered random nanostructures. The structural setup is the same as that in Fig. 2(d) of the main text. The validity of our numerical approach has been verified by the excellent agreement with results from Refs. [S9,S10].

➢ **Relation between the localization length and the system size needed for Brewster-anomaly-mediated free-electron radiation**

The system size required for delocalized, directional free-electron radiation can be quantitatively estimated from the localization length near the Brewster angle. Specicially, in Fig. S2(a), we define $\Delta\theta_{\rm loc}$ as the angular width within which the localization length exceeds the system sizes of interest. For the same system sizes, the radiation widths $\Delta\theta$ are extracted from free-electron radiation spectra shown in Figs. S2(b)–S2(d). As shown in Fig. S2(e), $\Delta\theta_{\rm loc}$ and $\Delta\theta$ are comparable; for example, $\Delta\theta_{\rm loc}\left(L/d_0 \approx 10^3\right) \le 10°$, while approximately $3\times10^3$ interfaces ($L/d_0 \approx 3\times10^3$) are sufficient to achieve directional free-electron radiation with $\Delta\theta(L'/d_0) \le 10°$.

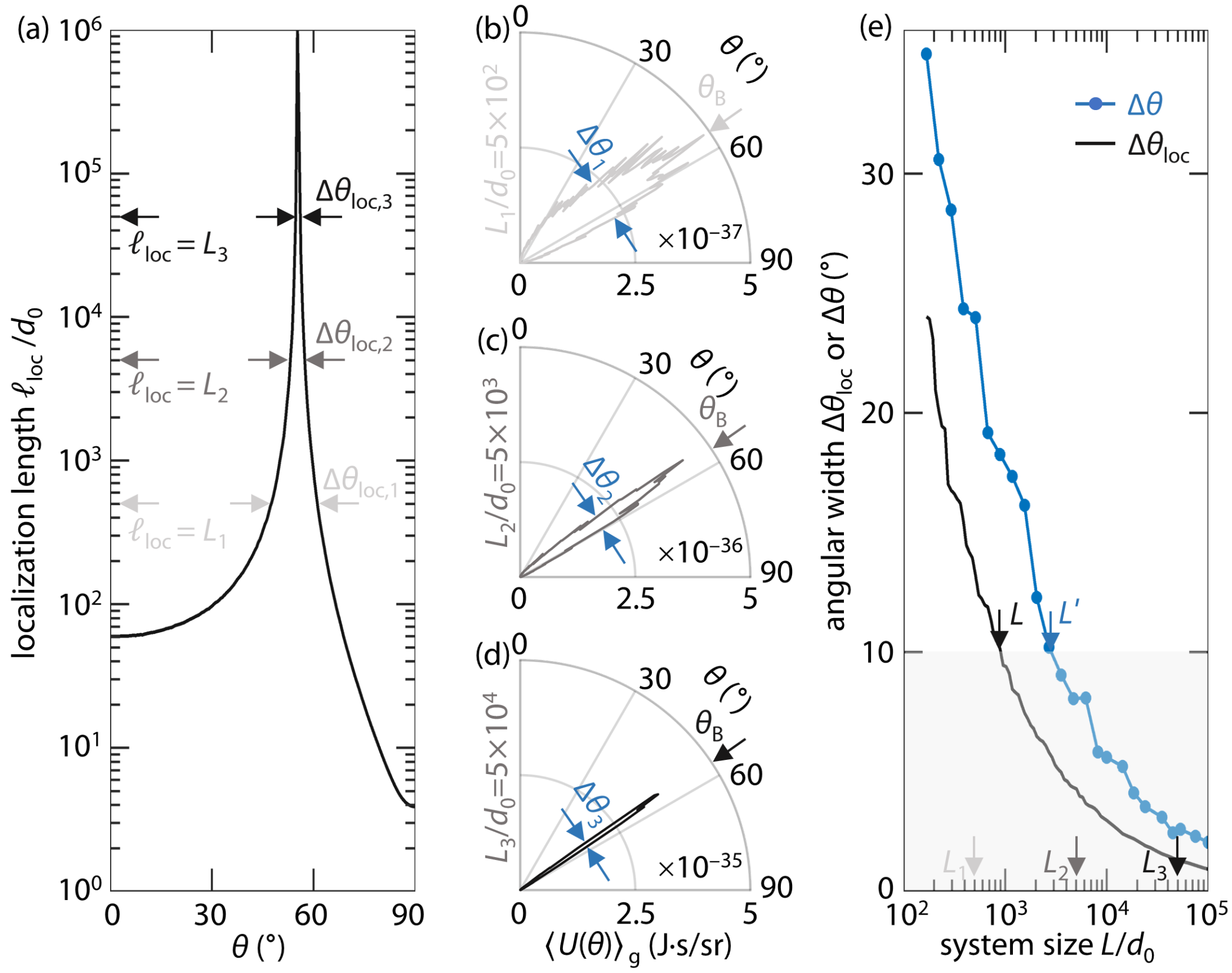

FIG. S2. Relation between the localization length and the system size needed for Brewster-anomaly-mediated free-electron radiation. (a) Dependence of the localization length on the radiation angle. (b)–(d) Configurationally averaged angular spectral energy densities for different system sizes. (e) Dependence of the angular widths on the system size. The structural setups are the same as those in Fig. 2 of the main text.

## S3 More discussion on Brewster-anomaly-mediated free-electron radiation

### S3.1 Robustness against variations in the disorder distribution

In this subsection, we show in Fig. S3 that intense and directional free-electron radiation persists for various disorder distributions.

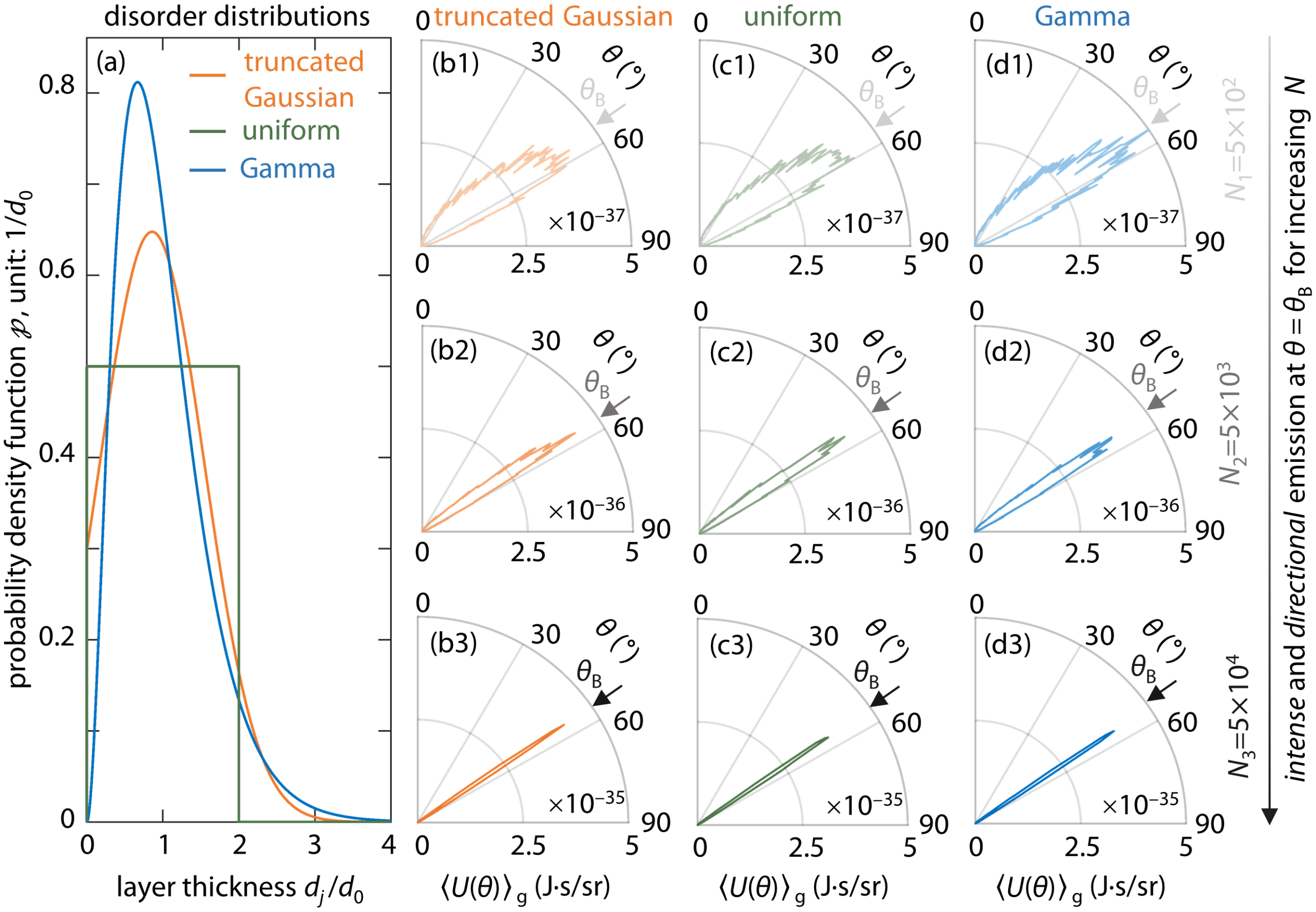

FIG. S3. Robustness of Brewster-anomaly-mediated free-electron radiation against variations in the disorder distribution. (a) Probability density functions of various disorder distributions. (b1)–(d3) Configurationally averaged angular spectral energy densities $\langle U(\theta) \rangle_g$ for various disorder distributions. For illustration, the structural setups are the same as those in Figs. 2(b)–2(d) of the main text, except that the layer thicknesses follow truncated Gaussian, positive uniform, and Gamma distributions with the same mean $d_0$ and standard deviation $\sigma = d_0/\sqrt{3}$ in (b1)–(b3), (c1)–(c3), and (d1)–(d3), respectively. Different rows correspond to different numbers of interfaces. The persistence of the intense and directional radiation for different probability density functions demonstrates that our finding is robust against the specific choice of disorder distribution.

## S3.2 Influence of variations in material permittivity

In this subsection, we study the influence of variations in material permittivity. Although the angular spectrum of a single realization may be noticeably affected, as shown in Figs. S4(a)–S4(c), these effects are averaged out by configurational averaging. In other words, free-electron radiation mediated by Brewster-anomaly delocalization is statistically robust against variations in material permittivity. Another origin of this robustness is that the Brewster angle itself can be relatively insensitive to permittivity variations, as shown in Fig. S4(d).

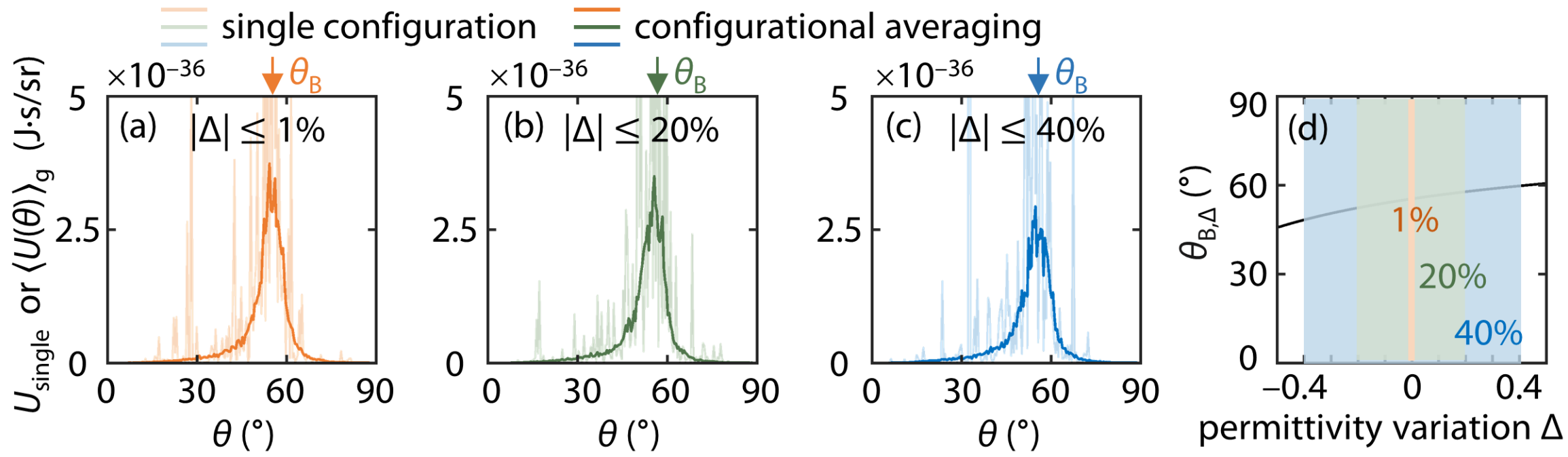

FIG. S4. Influence of variations in material permittivity on the delocalized free-electron radiation mediated by Brewster anomalies. (a)–(c) Single-realization and configurationally averaged angular spectral energy densities for different levels of permittivity variation. The structural setup is the same as that in Fig. 2(c) of the main text, except that the permittivity of dielectric II varies as $\varepsilon_{\mathrm{II},\Delta} = \varepsilon_{\mathrm{II}}(1+\Delta)$, where $\Delta$ is a uniform random variable and $\varepsilon_{\mathrm{II}} = 2.1$ is the unperturbed permittivity used in the main text. We set $|\Delta| \leq 1\%$, $|\Delta| \leq 20\%$, and $|\Delta| \leq 40\%$ in (a), (b), and (c), respectively. (d) Dependence of the Brewster angle on permittivity variations. Specifically, $\theta_{\mathrm{B},\Delta} = \arcsin\sqrt{\frac{\varepsilon_{\mathrm{II},\Delta}}{1+\varepsilon_{\mathrm{II},\Delta}}}$. Overall, Brewster-anomaly-mediated free-electron radiation is statistically robust against permittivity variation owing to the combined effects of configurational averaging and the relative insensitivity of the Brewster angle to such variations.

## S3.3 Influence of material loss

In this subsection, we study the influence of material loss. We show in Fig. S5 that free-electron radiation mediated by Brewster-anomaly delocalization exhibits a certain degree of tolerance to material loss. In particular, many low-loss dielectrics with loss tangents as low as $10^{-5}$ are available in practice [S12–S14]. In this regime, the intensity and directionality of the free-electron radiation remain essentially unaffected.

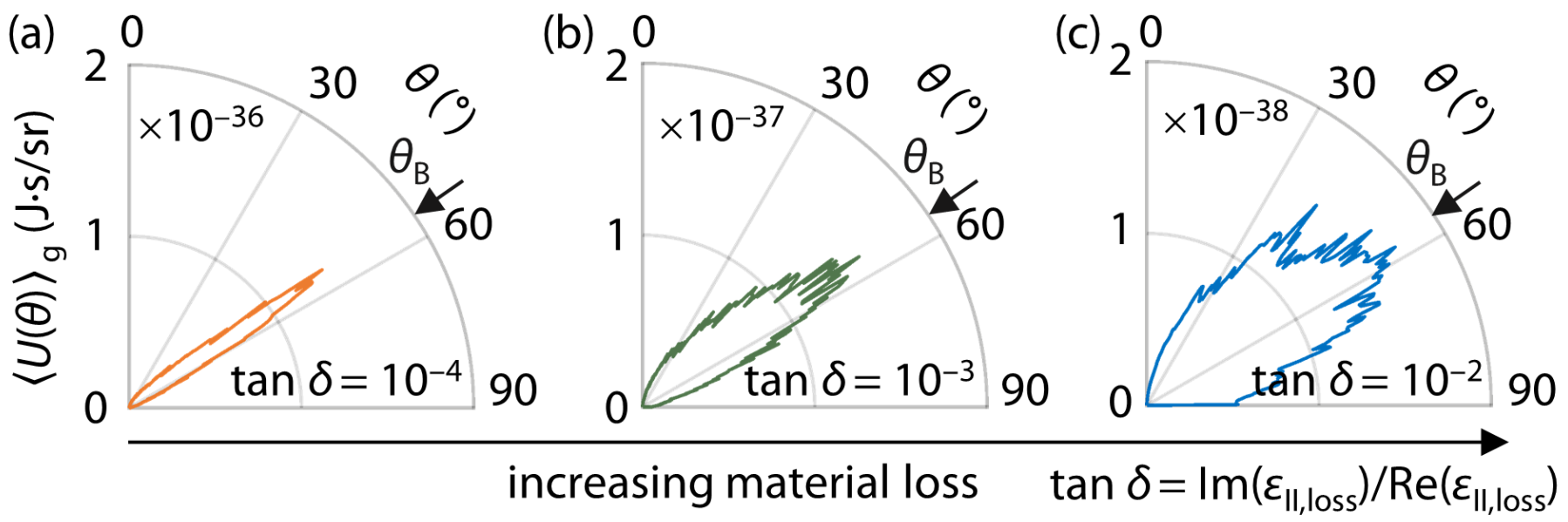

FIG. S5. Influence of material loss on the delocalized free-electron radiation mediated by Brewster anomalies. (a)–(c) Configurationally averaged angular spectral energy densities $\langle U(\theta)\rangle_{\mathrm{g}}$ for different levels of material loss. The structural setup here is the same as that in Fig. 2(c) of the main text, except that material loss is introduced, namely $\varepsilon_{\mathrm{II,loss}} = \varepsilon_{\mathrm{II}}(1 + i\tan\delta)$, where $\tan\delta = \mathrm{Im}(\varepsilon_{\mathrm{II,loss}})/\mathrm{Re}(\varepsilon_{\mathrm{II,loss}})$ is the loss tangent. For illustration, $\tan\delta$ is set to $10^{-4}$, $10^{-3}$, and $10^{-2}$ in (a)–(c), respectively.

### S3.4 Realistic material choice

In this subsection, we analyze realistic material choice. A necessary material condition is that radiation emitted at the Brewster angle of the layered medium can freely outcouple into the substrate and superstrate regions. Mathematically, this condition is written as $k_{\perp,\mathrm{B}} = k_0\sqrt{\varepsilon_{\mathrm{I}}\varepsilon_{\mathrm{II}}/(\varepsilon_{\mathrm{I}}+\varepsilon_{\mathrm{II}})} < k_0\sqrt{\varepsilon_{\mathrm{sur}}}$, where $k_{\perp,\mathrm{B}}$ is the wavevector component parallel to the interface at the Brewster angle and $k_0\sqrt{\varepsilon_{\mathrm{sur}}}$ is the wavevector in the substrate and superstrate regions. If $\varepsilon_{\mathrm{sur}} = 1$, one has $\theta_{\mathrm{B}} = \arcsin\sqrt{\varepsilon_{\mathrm{I}}\varepsilon_{\mathrm{II}}/(\varepsilon_{\mathrm{I}}+\varepsilon_{\mathrm{II}})} < 90°$. Figure S6(a) shows the permissible range of permittivity combinations satisfying this condition, which actually encompasses many commonly used dielectric materials. For example, in Fig. S6(b), we select realistic materials within this range while including their actual dispersion and loss. As a result, in Figs. S6(c) and S6(d), the free-electron radiation mediated by Brewster anomalies remains robust against realistic material dispersion and loss.

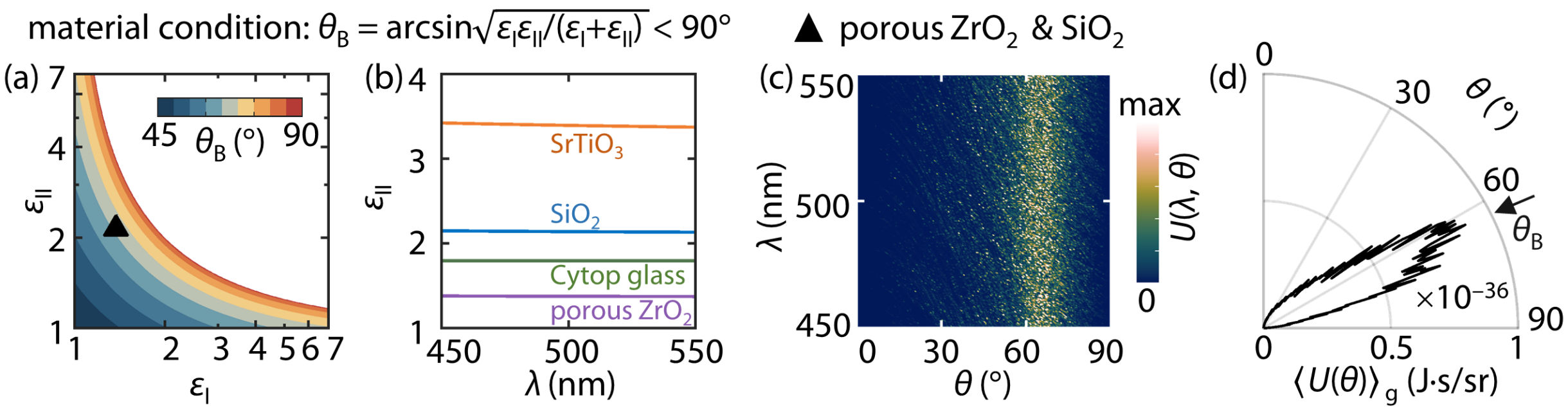

FIG. S6. Delocalized free-electron radiation mediated by Brewster anomalies from a layered nanostructure composed of realistic materials. (a) Brewster angle $\theta_{\mathrm{B}}$ mapped in the permittivity parameter space. (b) Wavelength dependence of the relative permittivities $\varepsilon_{\mathrm{r}}(\lambda)$ of some common dielectrics. Notably, low-loss dielectrics are widely available; the dielectrics exemplified in (b) exhibit loss tangents on the order of $10^{-5}$ in the visible spectrum [S12–S15], i.e., $\tan\delta = \mathrm{Im}(\varepsilon_{\mathrm{r}})/\mathrm{Re}(\varepsilon_{\mathrm{r}}) \leq 10^{-5}$ . (c),(d) Single-realization and configurationally averaged angular spectral energy densities for a layered random nanostructure composed of realistic materials. For illustration, $SiO_2$ [S12] and porous $ZrO_2$ [S13] are chosen and the loss tangent is set to $10^{-5}$. The other structural setups in (c) and (d) are the same as those in Figs. 4(c) and 2(c) of the main text, respectively.

## S3.5 Justification of the constant-velocity approximation

In this subsection, we highlight that our work is formulated within the constant-velocity approximation, meaning that the change in the electron velocity *caused by spontaneous photon emission* is safely neglected [S16]. The underlying reason is that the total energy of the emitted photons [see the energy spectrum in Fig. S7(a)] generally accounts for only a very small fraction of the electron's kinetic energy; for example, this fraction is as low as $10^{-6}$ under our structural setup [Fig. S7(b)]. On the other hand, unwanted inelastic electron scattering *caused by the electron directly impinging on the constituent dielectrics* can be substantially reduced in practice by drilling an aperture through the sample centered on the electron trajectory [S17–S19], as shown in Fig. S7(c). In this configuration, the evanescent field associated with the swift electron can still interact efficiently with the designed structure, while the electron passes through the sample with substantially suppressed inelastic scattering.

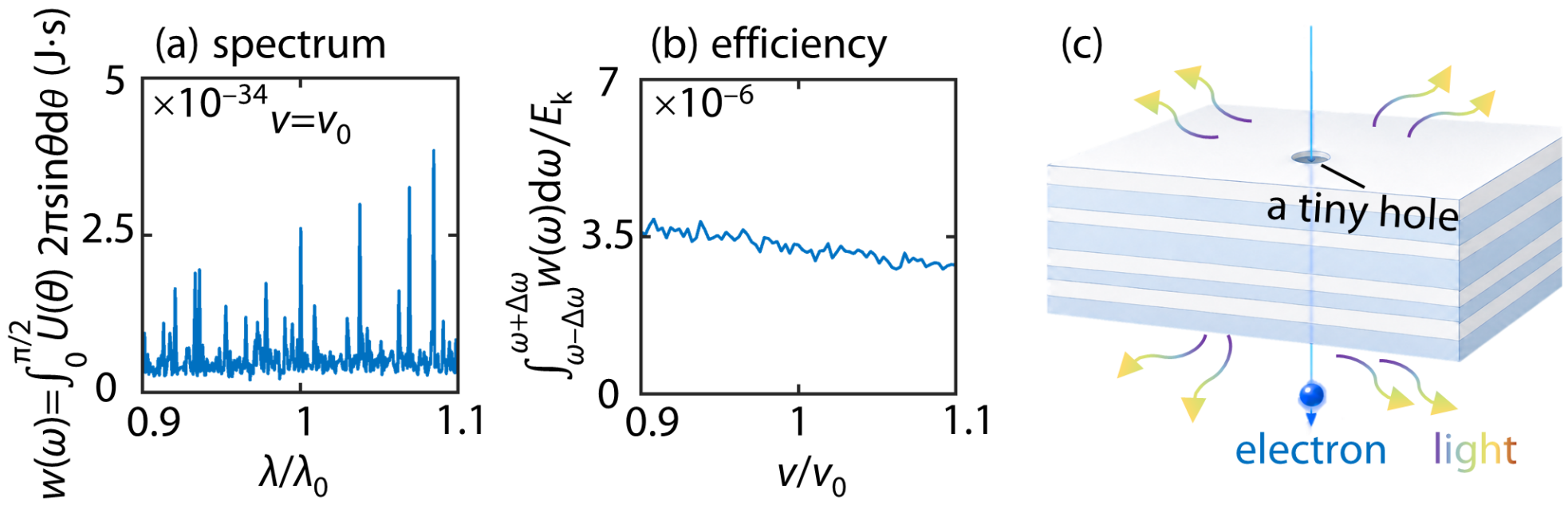

FIG. S7. Justification of the constant-velocity approximation. (a) Radiation frequency spectrum, $w(\omega) = \int_0^{\pi/2} U(\theta)\, 2\pi \sin\theta \,\mathrm{d}\theta$. (b) Radiation efficiency for various electron velocities, defined as the ratio of the total radiated energy to the electron kinetic energy, $\int_{\omega_0-\Delta\omega}^{\omega_0+\Delta\omega} w(\omega)\mathrm{d}\omega \,/E_{\mathrm{k}}$. The structural configurations are the same as those in Figs. 4(b) and 4(c) of the main text. (c) Schematic of a possible experimental realization. To allow electrons to pass safely through the layered structure, an aperture with a subwavelength or deeply subwavelength diameter can be introduced along the electron trajectory. Owing to the extremely low radiation efficiency and the effective suppression of the unwanted inelastic electron scattering, the electron velocity can be regarded as constant.